\pdfoutput=1
\documentclass[twocolumn,showpacs,prb,eqsecnum,citeautoscript,amsmath,amssymb,floatfix,superscriptaddress
]{revtex4-1}
\usepackage{amsmath,amssymb,color}
\usepackage{bm}
\usepackage{graphicx}
\usepackage[latin1]{inputenc}
\usepackage[T1]{fontenc}
\usepackage[FIGTOPCAP,nooneline]{subfigure}
\newcommand{\bd}{\bm}

\begin{document}

\title{Spontaneous ferromagnetism in the spinor Bose gas with
Rashba spin-orbit coupling}
 
\author{Kira Riedl}
\affiliation{Institut f\"{u}r Theoretische Physik, Universit\"{a}t
  Frankfurt,  Max-von-Laue Strasse 1, 60438 Frankfurt, Germany}

\author{Casper Drukier}
\affiliation{Institut f\"{u}r Theoretische Physik, Universit\"{a}t
  Frankfurt,  Max-von-Laue Strasse 1, 60438 Frankfurt, Germany}
  
\author{Peter Zalom}
\affiliation{Institut f\"{u}r Theoretische Physik, Universit\"{a}t
  Frankfurt,  Max-von-Laue Strasse 1, 60438 Frankfurt, Germany}
\affiliation{Institute of Experimental Physics, Slovak Academy of Sciences, Watsonova 47, 040 01 Ko\v{s}ice, Slovakia}
   
\author{Peter Kopietz}
  
\affiliation{Institut f\"{u}r Theoretische Physik, Universit\"{a}t
  Frankfurt,  Max-von-Laue Strasse 1, 60438 Frankfurt, Germany}

\date{April 18, 2013}

 \begin{abstract}
We show that in the two-component Bose gas with Rashba spin-orbit
coupling an arbitrarily small attractive interaction
between bosons with opposite spin
induces spontaneous ferromagnetism
below a finite critical temperature $T_c$.
In the ferromagnetic phase the single-particle spectrum exhibits a unique
minimum in momentum space in the direction of the magnetization.
For sufficiently small temperatures below $T_c$ 
the bosons eventually condense into the unique state at the bottom of the spectrum, 
forming a ferromagnetic Bose-Einstein condensate.

\end{abstract}

\pacs{67.85.Fg, 03.75.Mn, 05.30.Jp}

\maketitle

\section{Introduction}


Due to recent progress in the field of ultracold
gases, spinor Bose-Einstein
condensates with various types of spin-orbit coupling can now be realized experimentally \cite{Lin09a,Dalibard11} by using
spatially varying laser fields to couple internal pseudo-spin degrees of freedom to the 
momentum.  These experiments
have motivated many theoretical investigations of spin-orbit coupled multi-component Bose gases \cite{Stanescu08,Wang10,Gopalakrishnan11,Ho11,Ozawa11,Barnett12,Sedrakyan12,Hu12,Cui12,Liao12,Li12,Zhou13,Yu13}.
Of particular interest have been two-component bosons with isotropic
Rashba-type \cite{Rashba60}  spin-orbit coupling, where in the absence of interactions
the energy dispersion assumes a minimum on a circle
in momentum space \cite{Stanescu08}.
If the bosons do not condense,
such a surface in momentum space 
can be called a Bose surface, in analogy with the Fermi surface 
of an electronic system \cite{Paramekanti02,Sachdev02}. 
Of course, bosons do not obey the Pauli exclusion principle,
so that the  Bose surface cannot be the boundary between occupied and unoccupied states;
however, the Bose surface defines the location of the low-energy excitations
in the system, similar to the Fermi surface of an electronic system.

The fact that Bose-Einstein condensation (BEC) in Bose systems where 
the dispersion has  degenerate minima on a surface in momentum space
differs qualitatively from conventional BEC has been pointed out a long time ago by
Yukalov~\cite{Yukalov78}. He studied BEC
in an interacting Bose system whose
energy dispersion is  minimal on a sphere in momentum space.
Assuming that the bosons condense 
with equal weight into all states on this sphere,
he showed that the condensed state 
does not exhibit off-diagonal long-range order and is also not superfluid.

Due to the degeneracy
of the single-particle energy in the spinor Bose gas with Rashba-type spin-orbit coupling, 
in the non-interacting limit BEC is prohibited 
at any finite temperature. 
Hence, finite temperature 
BEC in this system must be an interaction effect \cite{Ozawa11,Hu12}.
Interactions are expected to remove the
ground state degeneracy and various types of exotic ground states have been 
proposed \cite{Stanescu08,Wang10,Gopalakrishnan11,Ho11,Ozawa11,Barnett12,Sedrakyan12,Hu12,Cui12,Liao12,Li12,Zhou13,Yu13}. 
Which phase is realized experimentally depends
on the specific properties of the interaction.
At this point a generally accepted agreement
on the nature of the ground state  in the spinor Bose gas
with Rashba-type  spin-orbit coupling has not been reached.

Phase transitions in systems whose fluctuation spectrum
exhibits minima on a surface in momentum space
form their own universality class, the so-called
Brazovskii universality class \cite{Brazovskii75}; 
for example, the critical behavior
in cholesteric liquid crystals belongs to this class \cite{Hohenberg95}.
Because scaling transformations and mode elimination in
renormalization group calculations should be defined relative to the
low-energy manifold, the classification of interaction
vertices in systems belonging to the Brazovskii universality class
is different from the corresponding classification in systems where the low-energy manifold consists of a single point;
in particular, all two-body scattering processes where the momenta of the
incoming and the outgoing particles lie on the low-energy manifold
are marginal, so that in renormalization group calculations one should keep track
of infinitely many marginal couplings. Note that also
normal fermions belong to the Brazovskii universality class, because
the Fermi surface can be identified with the low-energy manifold relative to which
scaling transformations should be defined \cite{Shiwa06,Kopietz01,Kopietz10}.
Two-component bosons with Rashba-type spin-orbit coupling
are therefore another example for  a quantum system which belongs
to the Brazovskii universality class.

In this work we shall further investigate interaction effects on
spinor Bose gases with Rashba-type spin-orbit coupling.
We shall consider  the specific case where
the interaction $g_{\bot}$ between two bosons with opposite pseudo-spin
is attractive.
We find that in this case for any finite density an arbitrarily small interaction
$g_{\bot} < 0$ leads to spontaneous ferromagnetism
below some finite temperature $T_c$.
In the ferromagnetic phase, the single-particle dispersion has
a unique minimum in momentum space, so that
the bosons eventually condense at some temperature $T_{\rm BEC} < T_c$ into the
unique single-particle state at the bottom of the spectrum.

\section{Spin-orbit coupled bosons}

We consider a
two-component Bose gas with Rashba
spin-orbit coupling and a two-body interaction which is
invariant under rotations around the $z$-axis in spin-space. The
Hamiltonian is given by ${\cal{H}}  =  {\cal{H}}_0+{\cal{H}}_{\rm int}$,
with
 \begin{eqnarray}
{\cal{H}}_0 & = & \sum_{\bd{k}} 
 ( a^{\dagger}_{\bd{k} \uparrow } , a^\dagger_{\bd{k} \downarrow} )
 \left[  \frac{\bd{k}^2  - 2 k_0  {\bd{k}}_{\bot}
 \cdot \bd{\sigma}  }{2m}
 \right] 
 \left( \begin{array}{c} a_{\bd{k} \uparrow}  \\ a_{\bd{k} \downarrow} 
 \end{array} \right), 
 \\
  {\cal{H}}_{\rm int}  & = & \frac{1}{ 2 V} \sum_{ 
 \bd{k}_1^{\prime} \bd{k}_2^{\prime} \bd{k}_{2} \bd{k}_1 }
 \sum_{\sigma_1 \sigma_2}
 \delta_{ \bd{k}_1^{\prime}+ \bd{k}_2^{\prime}, \bd{k}_{2}+ \bd{k}_1}
 \nonumber
 \\
 &   \times &
 U_{\sigma_1 \sigma_2} (   \bd{k}_1^{\prime},  \bd{k}_2^{\prime} ; \bd{k}_2 , \bd{k}_1 )
  a^{\dagger}_{ \bd{k}_1^{\prime} \sigma_1}
 a^{\dagger}_{ \bd{k}_2^{\prime} \sigma_2}
 a_{ \bd{k}_2 \sigma_2}
a_{ \bd{k}_1 \sigma_1},
 \hspace{7mm}
  \label{eq:H1}
 \end{eqnarray}
where $\bd{k}_{\bot} = k_x \hat{\bd{x}} + k_y \hat{\bd{y}}$
is the projection of the wave-vector $\bd{k}$ onto the $xy$-plane, and
the components of the vector operator
$\bd{\sigma}$  are the usual Pauli matrices. 
The wave-vector $k_0$ measures the strength of the spin-orbit coupling.
The invariance of the interaction under spin-rotations around the
$z$-axis implies that the function 
$U_{\sigma_1 \sigma_2} (   \bd{k}_1^{\prime},  \bd{k}_2^{\prime} ; \bd{k}_2 , \bd{k}_1 )$
is symmetric with respect to the simultaneous 
permutation of its incoming and outgoing labels\cite{Kopietz10}
 \begin{equation}
 U_{\sigma_1 \sigma_2} (   \bd{k}_1^{\prime},  \bd{k}_2^{\prime} ; \bd{k}_2 , \bd{k}_1 ) =
 U_{\sigma_2 \sigma_1} (   \bd{k}_2^{\prime},  \bd{k}_1^{\prime} ; \bd{k}_1 , \bd{k}_2 ).
 \end{equation}
For simplicity, we assume that the bare interaction is momentum independent,
so that the interaction is characterized by three different coupling constants
 \begin{equation}
 g_{\uparrow} = U_{ \uparrow  \uparrow} (0); \; \;
 g_{\downarrow} = U_{ \downarrow  \downarrow} (0); \; \;
 g_{\bot} = U_{ \uparrow  \downarrow} (0) = U_{ \downarrow  \uparrow} (0),
 \label{eq:gdef}
 \end{equation}
and ${\cal{H}}_{\rm int}$ simplifies to
 \begin{eqnarray}
 {\cal{H}}_{\rm int}
 & = & \frac{1}{ 2 V} \sum_{ 
 \bd{k}_1^{\prime} \bd{k}_2^{\prime} \bd{k}_{2} \bd{k}_1 }
 \delta_{ \bd{k}_1^{\prime}+ \bd{k}_2^{\prime}, \bd{k}_{2}+ \bd{k}_1   }
\Bigl[
 g_{\uparrow}
  a^{\dagger}_{ \bd{k}_1^{\prime} \uparrow}
 a^{\dagger}_{ \bd{k}_2^{\prime} \uparrow}
 a_{ \bd{k}_2 \uparrow}
a_{ \bd{k}_1 \uparrow}
 \nonumber
 \\ 
& & \hspace{-7mm}
+ g_{\downarrow}
  a^{\dagger}_{ \bd{k}_1^{\prime} \downarrow}
 a^{\dagger}_{ \bd{k}_2^{\prime} \downarrow}
 a_{ \bd{k}_2 \downarrow}
 a_{ \bd{k}_1 \downarrow}
+ 2 g_{\bot}  a^{\dagger}_{ \bd{k}_1^{\prime} \uparrow}
 a^{\dagger}_{ \bd{k}_2^{\prime} \downarrow}
 a_{ \bd{k}_2 \downarrow}
a_{ \bd{k}_1 \uparrow}
\Bigr].
 \label{eq:Hint}
 \end{eqnarray}
Introducing the usual field operators 
 \begin{equation}
 \hat{\psi}_{\sigma} ( \bd{r} ) = \frac{1}{\sqrt{V}} 
 \sum_{\bd{k}} e^{ i \bd{k} \cdot \bd{r} }
 a_{\bd{k} \sigma},
 \end{equation}
and the corresponding density operators $\hat{\rho}_{\sigma} ( \bd{r} )
 =  \hat{\psi}^{\dagger}_{\sigma} ( \bd{r} ) \hat{\psi}_{\sigma} ( \bd{r} )$
the interaction  can be written as follows,
 \begin{eqnarray}
 {\cal{H}}_{\rm int} & =  & \frac{1}{2} \int d^{D} r 
 \Bigl[ g_{\uparrow} : \hat{\rho}^2_{\uparrow} ( \bd{r} ) :
 + g_{\downarrow}  : \hat{\rho}^2_{\downarrow} ( \bd{r} ) :
 \nonumber
 \\
 &  & \hspace{13mm} +
  2 g_{\bot}   : \hat{\rho}_{\uparrow} ( \bd{r} )   \hat{\rho}_{\downarrow} ( \bd{r} ) :
 \Bigr] \; ,
 \label{eq:Hint1}
 \end{eqnarray}
where $: \ldots :$ denotes normal ordering. 
Assuming for simplicity  
$g_{\uparrow} = g_{\downarrow} = g_{\parallel}$, we may write the interaction as
 \begin{equation}
 {\cal{H}}_{\rm int} =   \frac{1}{2} \int d^{D} r 
 \Bigl[ g_{\rho} : \hat{\rho}^2 ( \bd{r} ) :  + g_{\sigma} : \hat{\sigma}^2 ( \bd{r} ) :
 \Bigr],
 \label{eq:Hint2}
 \end{equation}
where $\hat{\rho} = \hat{\rho}_{\uparrow} + \hat{\rho}_{\downarrow}$ and
$\hat{\sigma} = \hat{\rho}_{\uparrow} - \hat{\rho}_{\downarrow}$ represent the density and the
spin density, and the associated couplings are
 \begin{equation}
 g_{\rho} = \frac{1}{2} ( g_{\parallel} + g_{\bot} ), \; \; \; 
 g_\sigma = \frac{1}{2} ( g_{\parallel} - g_{\bot} ).
 \end{equation}
From Eq.~(\ref{eq:Hint2}) it is clear that for
$g_\sigma < 0 $ (i.e., $g_{\bot} > g_{\parallel}$) 
states with a finite spin-density are favored; an example is
the standing wave spin-striped state proposed in Ref.~[\onlinecite{Wang10}],
where the
bosons condense simultaneously into two momentum states
with opposite momenta on the low-energy manifold in momentum space.
On the other hand,  $g_{\sigma} > 0$ 
favors states with vanishing spin density,
such as  plane wave condensate where 
the bosons condense into a single momentum state,
or the charge striped states discussed in Refs.~[\onlinecite{Ho11,Li12}].
In this work, we focus on the case where $g_{\bot}$ is negative.
We show below that even for infinitesimally small
$ g_{\bot} < 0  $ the system exhibits spontaneous ferromagnetism
in the plane of the spin-orbit coupling at sufficiently low temperatures.

\begin{figure}[tb]    
  \centering
  \includegraphics[keepaspectratio,width=\linewidth]{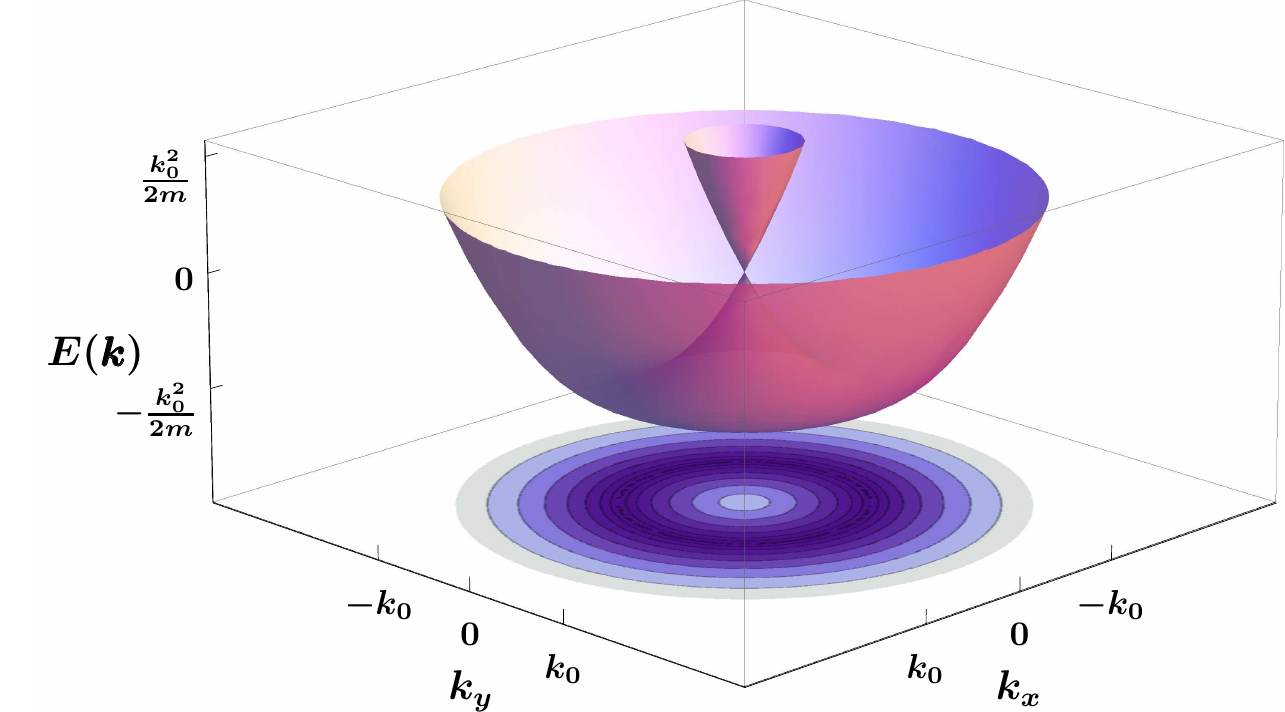}
  \caption{\label{fig:1}(Color online) %
Graph of the energy dispersions (\ref{eq:Ekpm}) of the spinor Bose gas with Rashba-type
spin-orbit coupling. The minimum of the lower helicity branch 
is a circle in the plane $k_z=0$ with radius $k_0$. The spacing between contours scales quartically. 
 \label{fig:dispersion}
} 
\end{figure}

To set up our notation, let us review the diagonalization of ${\cal{H}}_{0}$. Performing a momentum-dependent 
rotation in spin space around an axis $ \bd{\theta}_{\bd{k}} / | \bd{\theta}_{\bd{k}} |$ 
with angle $   | \bd{\theta}_{\bd{k}} |$,
 \begin{equation}
  \left( \begin{array}{c} a_{\bd{k} \uparrow}  \\ a_{\bd{k} \downarrow} 
 \end{array} \right) = 
e^{ - \frac{i}{2} {\bd{\sigma}} \cdot \bd{\theta}_{\bd{k}} }
\left( \begin{array}{c} a_{\bd{k} - }  \\ a_{\bd{k} +} 
 \end{array} \right),
 \label{eq:trafo}
\end{equation}
and using the fact that\cite{Rueckriegel12}
 \begin{equation}
e^{  \frac{i}{2} {\bd{\sigma}} \cdot \bd{\theta}_{\bd{k}} } \bd{\sigma}
e^{ - \frac{i}{2} {\bd{\sigma}} \cdot \bd{\theta}_{\bd{k}} } = e^{ \bd{\theta}_{\bd{k}} 
 \times } \bd{\sigma},
 \end{equation}
we obtain
\begin{equation}
 {\cal{H}}_0 = \sum_{\bd{k}} 
 ( a^{\dagger}_{\bd{k} - } , a^\dagger_{\bd{k} +} )
 \left[  \frac{\bd{k}^2  - 2 k_0  {\bd{k}}_{\bot}
 \cdot ( e^{ \bd{\theta}_{\bd{k}} \times } \bd{\sigma}) }{2m}
 \right] 
 \left( \begin{array}{c} a_{\bd{k} -}  \\ a_{\bd{k} +} 
 \end{array} \right) .
 \label{eq:H0diag1}
 \end{equation}
We now choose the rotation matrix $ e^{ \bd{\theta}_{\bd{k}} \times }$ such that
it rotates the z-axis into the direction $\hat{\bd{k}}_{\bot} =
 \bd{k}_{\bot} / | \bd{k}_{\bot} |$, i.e.,
 \begin{equation}
 \hat{\bd{k}}_{\bot} =  e^{ \bd{\theta}_{\bd{k}} \times } \hat{\bd{z}},
 \end{equation}
which can be achieved by setting
 \begin{equation}
 \bd{\theta}_{\bd{k}} = \frac{\pi}{2} \hat{\bd{z}} \times \hat{\bd{k}}_{\bot}.
 \end{equation}
Due to the rotational invariance of the scalar product, we may hence write
 \begin{equation}
 \hat{{\bd{k}}}_{\bot}
 \cdot ( e^{ \bd{\theta}_{\bd{k}} \times } \bd{\sigma}) =
(  e^{ \bd{\theta}_{\bd{k}} \times } \hat{\bd{z}} ) 
\cdot ( e^{ \bd{\theta}_{\bd{k}} \times } \bd{\sigma}) =   \hat{\bd{z}} 
\cdot  \bd{\sigma} =   \sigma^z.
 \end{equation}
Our rotation matrix in spin space is then explicitly given by
 \begin{eqnarray}
 e^{ - \frac{i}{2} {\bd{\sigma}} \cdot \bd{\theta}_{\bd{k}} }
 & = & \cos \left( \frac{ \theta_{\bd{k}} }{2} \right)
 \left( \begin{array}{cc} 1 & 0 \\ 0 & 1 \end{array} \right)
 - i  \sin \left( \frac{ \theta_{\bd{k}} }{2} \right) \bd{\sigma} \cdot \hat{\bd{\theta}}_{\bd{k}}
 \nonumber
 \\
 & = & \frac{1}{\sqrt{2}}
 \left[  \left( \begin{array}{cc} 1 & 0 \\ 0 & 1 \end{array} \right) - i \bd{\sigma}
 \cdot ( \hat{\bd{z}} \times \hat{\bd{k}}_{\bot} ) \right]
 \nonumber
 \\
 & = &  \frac{1}{\sqrt{2}}  \left( \begin{array}{cc} 1 & - \hat{k}_x + i \hat{k}_y \\ 
 \hat{k}_x + i \hat{k}_y & 1 \end{array} \right)
 \nonumber
 \\
 & = &  \frac{1}{\sqrt{2}}  \left( \begin{array}{cc} 1 & - e^{ - i \varphi_{\bd{k}} } \\ 
  e^{  i \varphi_{\bd{k}} } & 1 \end{array} \right),
 \end{eqnarray}
where in the last line we have set $ \hat{k}_x = \cos \varphi_{\bd{k}}$ and
 $ \hat{k}_y = \sin \varphi_{\bd{k}}$.
In the new basis (which we shall call the helicity basis)
the non-interacting part of the Hamiltonian is
 \begin{equation}
 {\cal{H}}_0 = \sum_{\bd{k}} \sum_{\lambda = \pm} E_{\bd{k}  \lambda} 
 a^{\dagger}_{\bd{k} \lambda} a_{\bd{k} \lambda},
 \end{equation}
with energy dispersions
 \begin{equation}
 E_{\bd{k} \lambda} =     \frac{ \bd{k}^2}{2m}  + \lambda  v_0 | \bd{k}_{\bot} |  =
  \frac{ (| \bd{k}_{\bot} | +  \lambda k_0)^2  + k_z^2 - k_0^2    }{2m}.
 \label{eq:Ekpm}
 \end{equation}
Here $v_0 = k_0/m$ and the helicity index  $\lambda = \pm$  labels the two branches
of the dispersion.
A graph of these dispersions is shown in Fig.~\ref{fig:dispersion}.
Note that the energy  $E_{\bd{k},-}$ assumes its  minimum $- k_0^2/(2m)$
 on a circle of radius $k_0$ in the $xy$-plane, while  $E_{\bd{k} , +}$
is non-negative and vanishes only at $\bd{k}=0$.
Due to the momentum-dependent rotation in spin-space, the interaction
vertices in the helicity basis acquire a momentum-dependence.
For completeness we give the properly symmetrized
expressions for these vertices in Appendix~A.  
For our mean-field calculation it is more convenient to work in the original
spin basis.

\section{Mean-field theory for transverse ferromagetism}

\subsection{Derivation of the mean-field equations}

\begin{figure}[tb]    
  \centering
  \includegraphics[keepaspectratio,width=\linewidth]{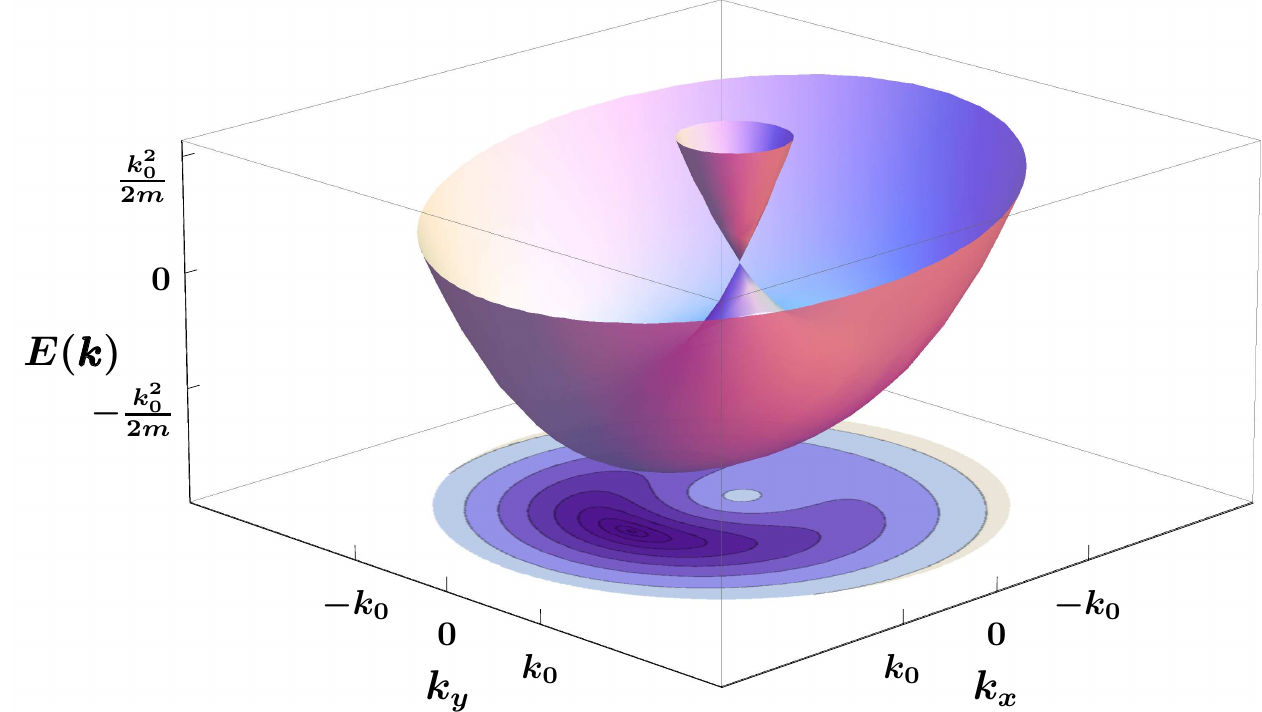}
  \caption{\label{fig:2}(Color online) %
Graph of the energy dispersions (\ref{eq:Ekpmp})
of the spinor Bose gas with Rashba-type spin-orbit coupling
and an effective magnetic field $\bd{h}_\bot - \bd{\Sigma} = v_0 \bd{p}$ pointing
in the direction of the positive $x$-axis. The lower helicity branch has a unique
minimum at
$\bd{k}_0 = k_0 \bd{p} / | \bd{p} |$. The spacing between contours scales quartically. 
} 
\label{fig:dispersionp}
\end{figure}

To study transverse ferromagnetism
we add a  uniform magnetic field
$\bd{h}_{\bot}$ in the $xy$-plane, so that
the non-interacting part of our Hamiltonian is now given by
 \begin{eqnarray}
 {\cal{H}}_0 & = & \sum_{\bd{k}} 
 ( a^{\dagger}_{\bd{k} \uparrow } , a^\dagger_{\bd{k} \downarrow} )
 \left[  \frac{\bd{k}^2  - 2 k_0  {\bd{k}}_{\bot}
 \cdot \bd{\sigma}  }{2m}
 - \bd{h}_{\bot} \cdot  \bd{\sigma}
 \right] 
 \left( \begin{array}{c} a_{\bd{k} \uparrow}  \\ a_{\bd{k} \downarrow} 
 \end{array} \right) .
 \nonumber
 \\
 & &
 \end{eqnarray}
For convenience we measure the magnetic field in units of energy.
The system exhibits
spontaneous ferromagnetism if the
magnetization remains finite when $\bd{h}_{\bot} \rightarrow 0$.
The spin-rotational invariance with respect to rotations around the
$z$-axis is then spontaneously broken.
While in the symmetric phase the self-energies
$\Sigma_{\sigma \sigma\prime}$ are diagonal in the spin-labels,
in the symmetry broken phase there are finite
off-diagonal components $\Sigma_{\uparrow \downarrow}$ and
$\Sigma_{\downarrow \uparrow}$.
Within the self-consistent Hartree-Fock approximation the self-energies
are  independent of momentum and frequency if
we start from a momentum-independent bare interaction.
The mean-field Hamiltonian is therefore of the form
 \begin{equation}
 {\cal{H}}_{\rm MF} = {\cal{H}}_0 + 
\sum_{\bd{k}} 
 ( a^{\dagger}_{\bd{k} \uparrow } , a^\dagger_{\bd{k} \downarrow} )
 \left( \begin{array}{cc} \Sigma_{\uparrow \uparrow} & \Sigma_{\uparrow \downarrow} \\
 \Sigma_{\downarrow \uparrow} & \Sigma_{\downarrow \downarrow} \end{array}
 \right) 
 \left( \begin{array}{c} a_{\bd{k} \uparrow}  \\ a_{\bd{k} \downarrow} 
 \end{array} \right) .
 \label{eq:Hmf}
 \end{equation}
Within the self-consistent Hartree-Fock approximation the self-energies are
 \begin{subequations}
 \begin{eqnarray}
 \Sigma_{\uparrow \uparrow} & = & 2 g_{\uparrow} \rho_{\uparrow} + 
 g_{\bot} \rho_{\downarrow},
 \\
\Sigma_{\downarrow \downarrow} & = & 2 g_{\downarrow} \rho_{\downarrow} + 
 g_{\bot} \rho_{\uparrow},
 \\
  \Sigma_{\uparrow \downarrow} & = &  g_{\bot}  \rho_{ \downarrow \uparrow},
 \\
 \Sigma_{\downarrow \uparrow} & = &  g_{\bot} {\rho}_{\uparrow \downarrow},
 \end{eqnarray}
\end{subequations}
where we have introduced the densities
 \begin{eqnarray}
 \rho_{\sigma} & = & \frac{1}{V} \sum_{\bd{k}} \langle a^{\dagger}_{\bd{k} \sigma} 
 a_{\bd{k} \sigma}  \rangle,
 \label{eq:rhodiagdef}
 \\
 \rho_{\uparrow \downarrow}  & = & \rho_{\downarrow \uparrow}^{\ast} = \frac{1}{V} \sum_{\bd{k}} 
 \langle a^{\dagger}_{\bd{k} \uparrow} 
 a_{\bd{k} \downarrow}  \rangle.
 \label{eq:rhooffdiagdef}
 \end{eqnarray}
Here the expectation values should be evaluated with the
grand canonical density matrix associated with the mean-field Hamiltonian
(\ref{eq:Hmf}). Note that our mean-field decoupling excludes states with broken translational invariance. This will be 
justified a posteriori from the fact that the irreducible ferromagnetic susceptibility is exponentially large at low 
temperatures [see Eq.~(\ref{eq:chilow})], such that, at least at weak coupling, the ferromagnetic instability is dominant.
Keeping in mind that $\rho_{\uparrow \downarrow} = M_x + i M_y$
can be expressed in terms of the components of 
the transverse magnetization 
$\bd{M}_{\bot} = M_x \hat{\bd{x}} + M_y \hat{\bd{y}}$,
we see that
the Cartesian components of the off-diagonal self-energies
are proportional to the corresponding components
of the magnetization,
 \begin{eqnarray}
 \Sigma_x & = &  \frac{1}{2} ( \Sigma_{\uparrow \downarrow} +  \Sigma_{\downarrow \uparrow} )
 = g_{\bot} M_x ,
 \label{eq:sigmaxdef}
 \\
 \Sigma_y & = &  \frac{i}{2} ( \Sigma_{\uparrow \downarrow} -  \Sigma_{\downarrow \uparrow} )
 = g_{\bot} M_y.
 \label{eq:sigmaydef}
  \end{eqnarray}
It is convenient to define in addition the self-energies
 \begin{eqnarray} 
 \Sigma_z & = &  \frac{1}{2} ( \Sigma_{\uparrow \uparrow} -  \Sigma_{\downarrow \downarrow} )
= g_{\uparrow} \rho_{\uparrow} - g_{\downarrow} \rho_{\downarrow} - 
 \frac{g_{\bot}}{2} ( \rho_{\uparrow} - \rho_{\downarrow} ) ,
 \\
 \Sigma_0 & = &  \frac{1}{2} ( \Sigma_{\uparrow \uparrow} +  \Sigma_{\downarrow \downarrow} )
= g_{\uparrow} \rho_{\uparrow} + g_{\downarrow} \rho_{\downarrow} + 
 \frac{g_{\bot}}{2} ( \rho_{\uparrow} + \rho_{\downarrow} ) ,
 \hspace{7mm}
\end{eqnarray}
and the wave-vector
 \begin{equation}
\bd{p}  =  (\bd{h}_{\bot} - \bd{\Sigma})/v_0,
 \label{eq:pdef}
 \end{equation}
where  $\bd{\Sigma} = \Sigma_x \hat{\bd{x}} +
 \Sigma_y \hat{\bd{y}} + \Sigma_z \hat{\bd{z}} $ 
is proportional to the internal magnetic field induced by the
interaction. 
The Hamiltonian can now be diagonalized via a momentum-dependent
rotation in spin space of the form (\ref{eq:trafo}).
The rotation matrix can be written as
 \begin{equation}
e^{ - \frac{i}{2} {\bd{\sigma}} \cdot \bd{\theta}_{\bd{k}} }
 =  \left( \begin{array}{cc} \cos ( \theta_{\bd{k}} / 2 )  & 
 - \sin ( \theta_{\bd{k}} / 2 ) e^{ - i \varphi_{\bd{k}}} \\
 \sin ( \theta_{\bd{k}} / 2 ) e^{ i \varphi_{\bd{k}}}
 &  \cos ( \theta_{\bd{k}} / 2 ) \end{array} \right),
 \end{equation} 
where
the rotation angles in the presence of
an effective magnetic field $\bd{h}_\bot - \bd{\Sigma} = v_0 \bd{p}$ are  now given by
 \begin{subequations}
 \begin{eqnarray}
 \cos \theta_{\bd{k}} & = &  \frac{p_z}{ | \bd{k}_{\bot} +  \bd{p} | },
 \\
  \sin \theta_{\bd{k}} & = & \sqrt{ 1 -  \frac{p_z^2}{ | \bd{k}_{\bot}+  \bd{p}| ^2  }},
 \\
 \cos \varphi_{\bd{k}} & = & \frac{ k_x + p_{ x} }{    \sqrt{ (k_x + p_ x)^2 + (k_y + p_y)^2 } } ,
 \\
  \sin \varphi_{\bd{k}} & = & \frac{ k_y + p_{ y} }{    \sqrt{ (k_x + p_ x)^2 + (k_y + p_y)^2 } } .
 \end{eqnarray}
 \end{subequations}
The energy dispersions of the eigenmodes are
 \begin{equation}
  E_{\bd{k} \lambda } = 
 \frac{ \bd{k}^2}{2m} + \Sigma_0 + \lambda v_0 | \bd{k}_{\bot} + \bd{p} |,
 \label{eq:Ekpmp}
 \end{equation}
where $\lambda = \pm$ labels again the helicity of the modes.
These dispersions are shown graphically in Fig.~\ref{fig:dispersionp}.
Obviously, for any finite $\bd{p}$ the degeneracy 
of the corresponding dispersion without magnetic field
shown in Fig.~\ref{fig:dispersion}
is completely removed, so that
  $E_{\bd{k} ,- }$ now has a unique minimum at
$\bd{k}_0 = k_0 \bd{p} / | \bd{p} |$.

To derive a self-consistency equation for the transverse magnetization, we simply
evaluate the off-diagonal density (\ref{eq:rhooffdiagdef}), using
the grand canonical density matrix associated with the
mean-field Hamiltonian on the right-hand side.
We thus obtain
\begin{eqnarray}
 \rho_{\uparrow \downarrow} &  = &   M_x + i M_y 
  =  -  \frac{1}{2 V} \sum_{\bd{k}  \lambda } \lambda  \sin \theta_{\bd{k}} e^{ i \varphi_{\bd{k}}}
n_{\bd{k} \lambda} ,
 \hspace{7mm}
 \label{eq:rhoupdown}
 \end{eqnarray}
where
 \begin{equation}
 n_{\bd{k} \lambda} = \frac{1}{ e^{ \beta ( E_{\bd{k} \lambda} - \mu ) } -1 }
 \end{equation} 
is the average occupation of the mode with energy $E_{\bd{k} \lambda }$.
Below we shall work at constant density, so that
we should eliminate the chemical potential $\mu$ in favor
of $\rho$. Therefore we need the diagonal densities 
(\ref{eq:rhodiagdef}),
 \begin{eqnarray}
 \rho_{\sigma}
 & = & \frac{1}{2V}
 \sum_{\bd{k}}  \sum_{\lambda} \left[ 1 - \sigma \lambda \frac{p_z}{ | \bd{k}_{\bot} + \bd{p} | }  
   \right] n_{\bd{k} \lambda},
 \end{eqnarray}
implying that the total density is
  \begin{equation}
 \rho = \sum_{\sigma} \rho_{\sigma} = \frac{1}{V} \sum_{\bd{k} , \lambda} n_{\bd{k}  \lambda}.
 \label{eq:rhoequation}
  \end{equation}
To show that the phase transition to the
ferromagnetic state is continuous,
it is useful to calculate
the grand canonical potential, which in a mean-field approximation
is given by
 \begin{equation}
 {\Omega}  ( T , \mu , h )
= T \sum_{\bd{k}} \sum_{\lambda = \pm}
 \ln \left[ 1 - e^{ - \beta ( E_{\bd{k} \lambda} - \mu )} \right] - \langle {\cal{H}}_{\rm int} \rangle,
 \label{eq:Omegaequation}
 \end{equation}
where the expectation value of the interaction part of the Hamiltonian is
 \begin{eqnarray}
 \langle {\cal{H}}_{\rm int} \rangle / V & = &  
 g_{\uparrow} \rho_{\uparrow}^2 + g_{\downarrow} \rho_{\downarrow}^2 
 + g_{\bot} ( \rho_{\uparrow} \rho_{\downarrow} + \bd{M}_{\bot}^2 )
 \nonumber
 \\
 & = & \left( \frac{g_{\parallel}}{2} + \frac{g_{\bot}}{4} \right) \rho^2
 +   g_{\bot} \bd{M}_{\bot}^2 ,
 \end{eqnarray}
and the second line holds for $g_{\uparrow} = g_{\downarrow}
 = g_{\parallel}$.
For simplicity, let us now choose $\bd{h}_\bot = h \hat{\bd{x}}$
so that $\bd{M}_{\bot} = M \hat{\bd{x}}$.
To explore the possibility of spontaneous transverse magnetization
at constant density, we should consider the Gibbs potential
 \begin{eqnarray}
 G ( T , \rho , M )
 & = & \Omega ( T , \mu  , {h}  )
+ V ( \mu \rho + {h} {{M}} ),
 \end{eqnarray}
where $\mu = \mu ( \rho, M )$ and $h = h ( \rho , M)$ 
should be determined by inverting the equations
 \begin{eqnarray}
 \rho & = &  - \frac{1}{V} \frac{ \partial \Omega }{\partial \mu }, \; \; \; 
 M  =   - \frac{1}{V} \frac{ \partial \Omega }{\partial h }.
 \end{eqnarray}

To study spontaneous transverse ferromagnetism, we shall later 
take the limit $h \rightarrow 0$. For simplicity,
we assume that
$g_{\uparrow} = g_{\downarrow} = g_{\parallel}$, so that $\rho_{\uparrow} = \rho_{\downarrow}
 = \rho/2$ and
$\Sigma_0 = ( g_{\parallel} + g_{\bot}/2 ) \rho$. As a consequence $\Sigma_z =0$
and $p_z =0$. Our self-consistency equation (\ref{eq:rhoupdown}) then 
reduces to the following
equation for the  transverse magnetization,
 \begin{equation}
  \bd{M}_{\bot}  = - \frac{1}{2 V} \sum_{\bd{k} \lambda} \lambda \frac{  {\bd{k}}_{\bot} + \bd{p} }{ 
 |  \bd{k}_{\bot} + \bd{p}  | } n_{\bd{k} \lambda }.
 \label{eq:mself1}
 \end{equation}
Note that $\bd{p}$ in the right-hand side depends again on
$\bd{M}_{\bot}$ 
via Eqs.~(\ref{eq:pdef}) and (\ref{eq:sigmaxdef},\ref{eq:sigmaydef});
for $g_{\uparrow} = g_{\downarrow}$  and $\bd{h} \rightarrow 0$
the relation between $\bd{p}$ and $\bd{M}_{\bot}$ is simply
$\bd{p} = - g_\bot \bd{M}_{\bot} / v_0$.
To determine the order of the phase transition for $h \rightarrow 0$, it is sufficient to consider the 
change in the free energy $F ( T , \rho  )  = \Omega + V \mu \rho$
for arbitrary magnetization $M$,
 \begin{equation}
 \Delta F ( T , \rho \;  M  )   = F ( T ,  \rho , h=0)_{M \neq 0}
  -  F( T , \rho ,  h=0)_{M=0}.
 \end{equation}
The physical state of the system at vanishing external field is determined by
 $\partial \Delta F ( T , \rho ; M ) / \partial M = 0$, 
which is another way of deriving
the  self-consistency equation  (\ref{eq:mself1}) for the order parameter $M$.

\subsection{Spectral densities}

\begin{figure}[tb]    
  \centering
  \subfigure[]{\includegraphics[width=1 \linewidth]{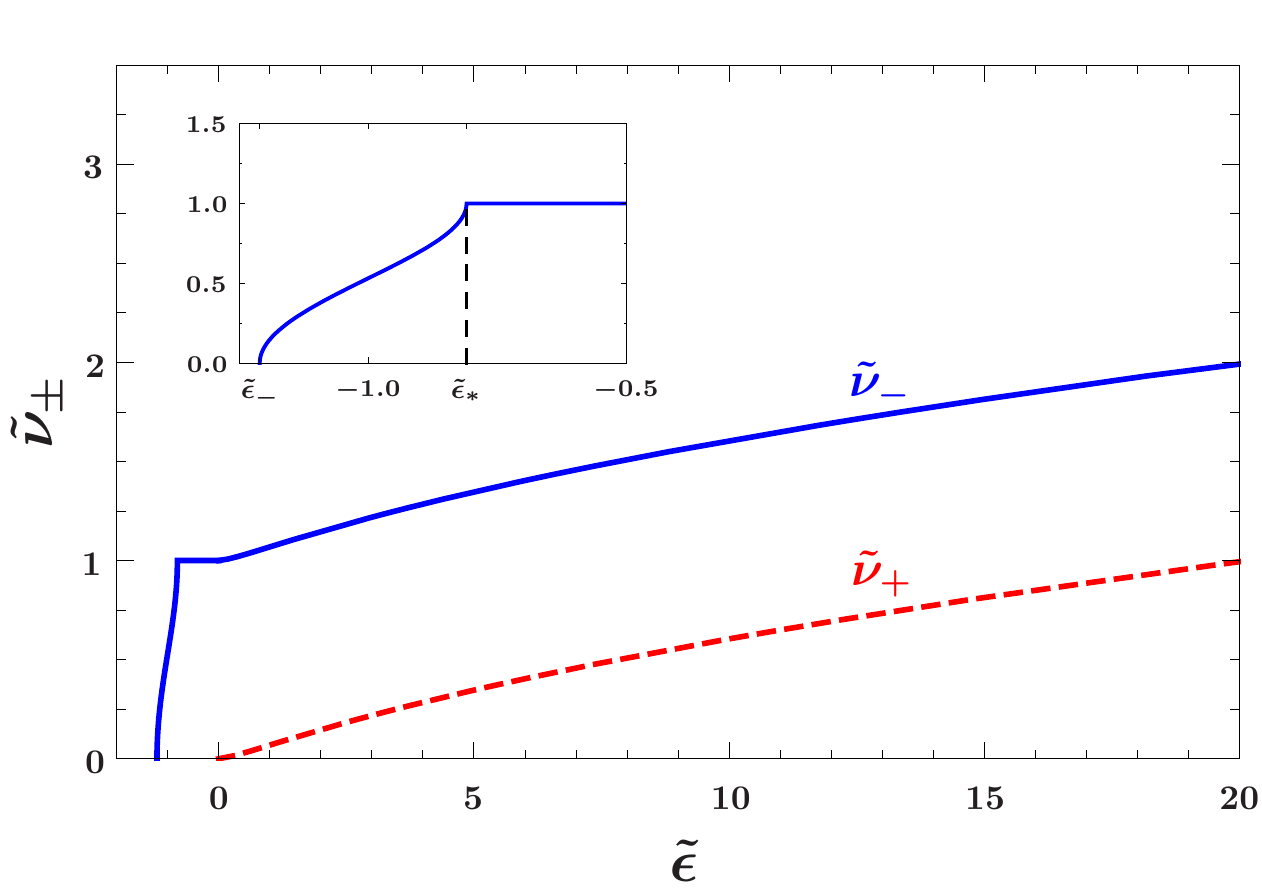}}
  \subfigure[]{\includegraphics[width=1 \linewidth]{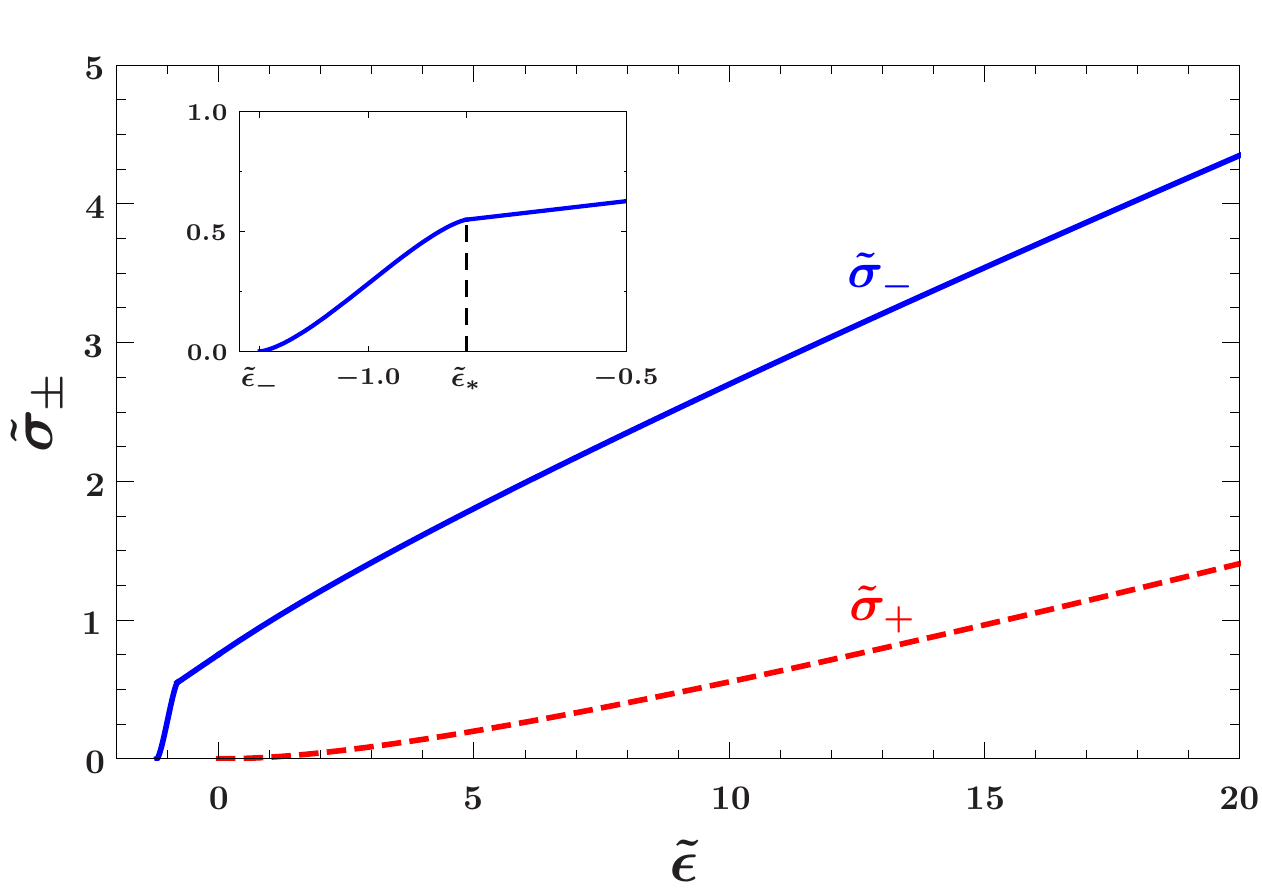}}
  \caption{\label{fig:3}(Color online) %
(a) Graph of the scaling functions $\tilde{\nu}_\pm ( \tilde{\epsilon} , \tilde{p} )$
of the density of states  defined via Eq.~(\ref{eq:nuscale})   for $\tilde{p} =0.1$,
see Eqs.~(\ref{eq:nuscale1int}, \ref{eq:nunegativexact}). (b) Graph of the scaling 
functions $\tilde{\sigma}_\pm ( \tilde{\epsilon} , \tilde{p} )$
of the weighted density of states  defined via Eq.~(\ref{eq:sigmascale})   for $\tilde{p} =0.1$,
see Eqs.~(\ref{eq:sigmascale1int}, \ref{eq:sigmanegativexact}). The solid lines correspond to the scaling functions in the $\lambda=-1$ branch while the dashed lines are the scaling functions in the $\lambda=1$ branch. The insets show a closeup of the negative energy part of the spectral functions in the lower helicity branch. We use $\tilde{\epsilon}_-= \left(\epsilon_--\Sigma_0-\left(p^2/2m\right)\right)/\epsilon_0$ and $\tilde{\epsilon}_\ast=\left(\epsilon_\ast-\Sigma_0-\left(p^2/2m\right)\right)/\epsilon_0$.
}
 \label{fig:scalingfuncs}
\end{figure}

\begin{figure}[tb]    
  \centering
  \subfigure[]{\includegraphics[width=0.4\linewidth]{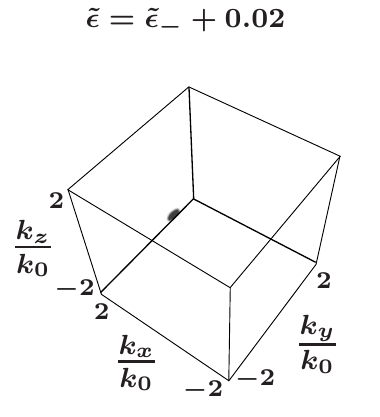}} \qquad
  \subfigure[]{\includegraphics[width=0.4\linewidth]{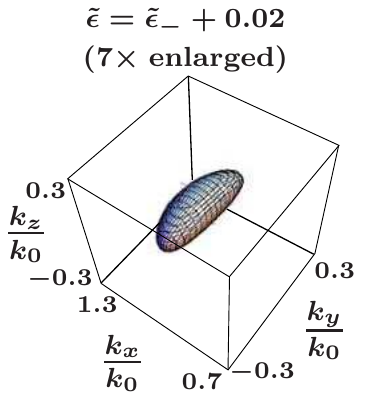}} \\
  \subfigure[]{\includegraphics[width=\linewidth]{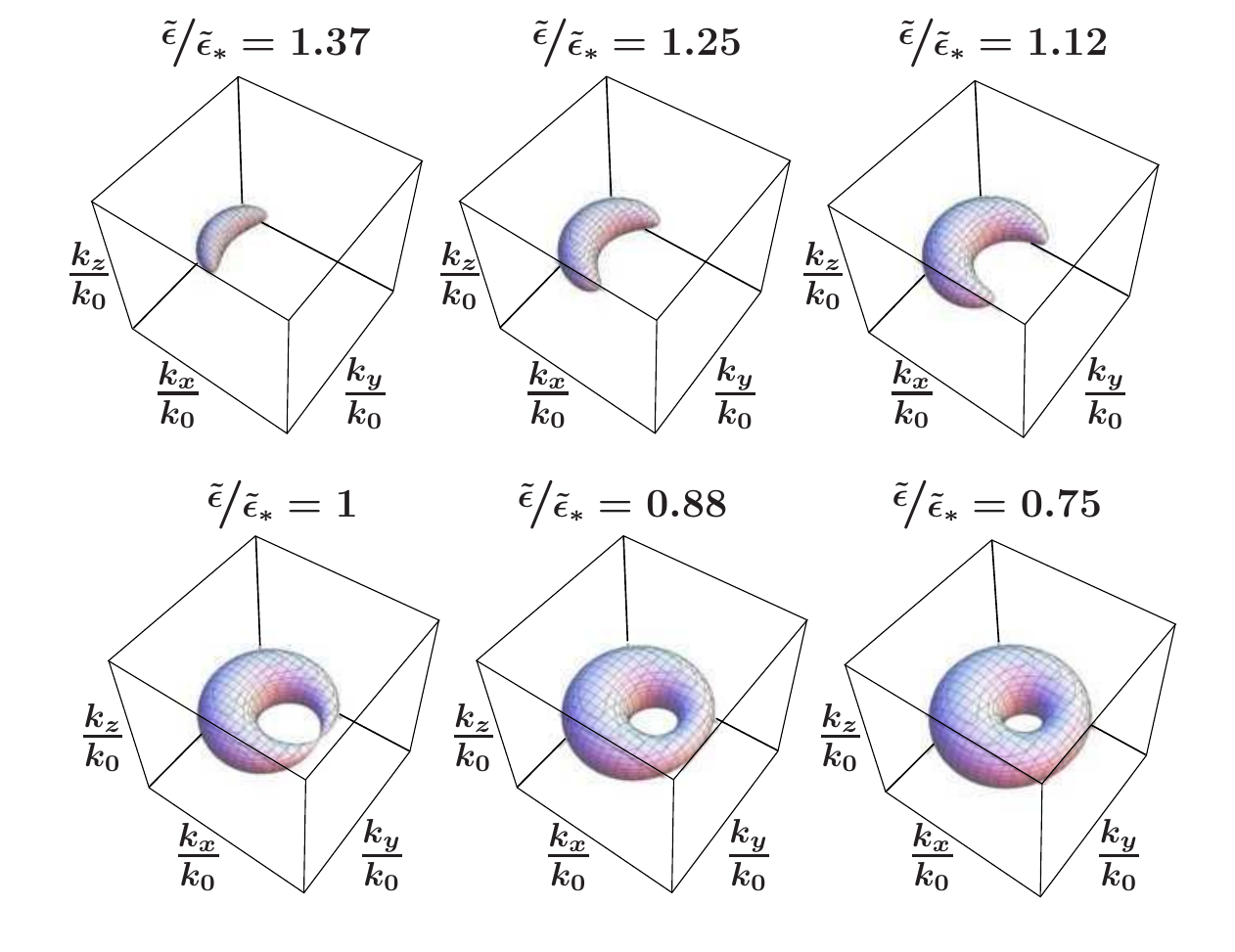}}
  \caption{\label{fig:energysurf}(Color online) %
(a) Surface of constant energy  $E_{\bd{k} , - } = \epsilon$ of the lower helicity branch in the ferromagnetic phase just above the lower energy threshold $\tilde{\epsilon}_-=-(1+\tilde{p})^2$. (b) Same as (a) but enlarged by a factor of seven. (c) Evolution of the constant energy surface for different energies. The scales are the same as in (a). All plots are for $\tilde{p}=0.1$. 
} 
 \label{fig:energysurface}
\end{figure}

To evaluate the integrals appearing in Eqs.~(\ref{eq:rhoequation})
and (\ref{eq:Omegaequation}), it is useful to introduce
the density of states
  \begin{equation}
 \nu_{\lambda} ( \epsilon, p  ) = \frac{1}{V} \sum_{\bd{k}} \delta ( \epsilon - E_{\bd{k} \lambda} )
 = \frac{1}{V} \sum_{\bd{k}^{\prime}} \delta ( \epsilon - E_{\bd{k}^{\prime} - \bd{p},  \lambda} ) ,
 \label{eq:dosdef}
 \end{equation}
where $p = ( h - g_{\bot} M)  / v_0$
and we have shifted $\bd{k}^\prime=\bd{k}+\bd{p}$ on the right-hand side.
The density equation (\ref{eq:rhoequation}) can then be written as
 \begin{equation}
 \rho  =   \int_{- \infty}^{\infty} d \epsilon \sum_{\lambda} 
 \nu_{\lambda} ( \epsilon, p ) \frac{1}{e^{ \beta ( \epsilon - \mu ) } -1 },
 \label{eq:densityeq}
 \end{equation}
while our expression (\ref{eq:Omegaequation}) for the grand canonical
potential per volume becomes
 \begin{eqnarray}
 \frac{\Omega ( T , \mu , h )}{V} & = &  
T \int_{- \infty}^{\infty} d \epsilon \sum_{\lambda}
 \nu_{\lambda} ( \epsilon, p ) \ln [ 1 - e^{ - \beta ( \epsilon - \mu ) } ]
 \nonumber
 \\
 & - & \frac{ g_{\parallel}}{2} \rho^2 -  g_{\bot} \left( \frac{\rho^2}{4} + M^2 \right).
 \end{eqnarray} 
It is also useful to rewrite the self-consistency equation for
$\bd{M}_{\bot} = M \hat{\bd{x}}$ in terms of
a generalized susceptibility as follows.
Shifting $\bd{k}^{\prime} = \bd{k} + \bd{p}$ in Eq.~(\ref{eq:mself1}) we obtain
 \begin{eqnarray}
 M & = &   - \frac{1}{2 V} \sum_{\bd{k}^{\prime} \lambda} \lambda 
 \frac{ k_x^{\prime}}{ | \bd{k}_{\bot}^{\prime} | } n_{ \bd{k}^{\prime} - \bd{p}  ,  \lambda}.
 \label{eq:mself2}
 \end{eqnarray}
We now introduce cylindrical coordinates in $\bd{k}^{\prime}$-space
and perform a partial integration
in the angular part. Rearranging terms we find that
the self-consistency equation (\ref{eq:mself2}) can be written as
\begin{equation}
 \frac{h}{M} = \frac{1}{\chi_{\bot} ( M ) } + g_{\bot},
 \label{eq:mselfnew}
 \end{equation}
where the irreducible susceptibility $\chi_{\bot} ( M )$ is defined by 
 \begin{equation}
 \chi_\bot ( M ) = - \frac{\beta}{2 V} \sum_{\bd{k}^{\prime} \lambda} 
\frac{ k^{\prime 2}_y}{ | \bd{k}^{\prime}_{\bot} | k_0 } \lambda 
 n_{ \bd{k}^{\prime} - \bd{p}  ,  \lambda} [ n_{ \bd{k}^{\prime} - \bd{p}  ,  \lambda} + 1].
 \end{equation}
It is easy to see that $\chi_{\bot} ( M ) > 0$, implying that, 
only for $g_{\bot} < 0$, the magnetization  $M$ can remain finite for $h \rightarrow 0$. 
From now on we shall therefore assume that $g_{\bot}$ is negative.
In this case the magnetization in the broken symmetry phase
satisfies the self-consistency equation
 \begin{equation}
 \chi_{\bot} ( M) = - \frac{1}{g_{\bot}}.
  \label{eq:selfchi}
 \end{equation}
Note that the physical susceptibility $M/h$ diverges at the critical point.
To evaluate $\chi_{\bot} ( M )$, we introduce the weighted density of states,
 \begin{eqnarray}
 {\sigma}_{\lambda} ( \epsilon, p  )
 & = &
\frac{1}{V} \sum_{\bd{k}^{\prime}}     \frac{ k_y^{\prime 2}}{ | \bd{k}_{\bot}^{\prime} | k_0}
 \delta ( \epsilon - E_{\bd{k}^{\prime} - \bd{p},  \lambda} ) .
 \label{eq:nutildedef}
 \end{eqnarray}
Then we may write
 \begin{eqnarray}
 \chi_{\bot} ( M )  & = &  - \frac{\beta}{2} \int_{- \infty}^{\infty} d \epsilon \sum_{\lambda} 
 \lambda \sigma_{\lambda} ( \epsilon, p ) \frac{
 e^{ \beta ( \epsilon - \mu ) }   }{[ e^{ \beta ( \epsilon - \mu ) } -1 ]^2 }.
 \hspace{7mm}
 \end{eqnarray}
Introducing cylindrical coordinates in  $\bd{k}^{\prime}$-space
in the above integrals defining
$\nu_{\lambda} ( \epsilon, p )$ and
$\sigma_{\lambda} ( \epsilon, p )$,
the integrations over $k_z^{\prime}$ and $| \bd{k}_{\bot}^{\prime} | $
can be carried out exactly so that 
we can write these functions as one-dimensional angular integrals.
The results can be written in the scaling form
 \begin{eqnarray}
 \frac{\nu_{\lambda} ( \epsilon, p )}{\nu_0} & = & 
 \tilde{\nu}_{\lambda}
 \left( \frac{ \epsilon  - \Sigma_0  - \frac{ p^2}{2m} }{\epsilon_0},
 \frac{p}{k_0} \right),
 \label{eq:nuscale}
 \\
  \frac{\sigma_{\lambda} ( \epsilon, p )}{\nu_0} & = & 
 \tilde{\sigma}_{\lambda}
 \left( \frac{ \epsilon  - \Sigma_0  - \frac{ p^2}{2m} }{\epsilon_0},
 \frac{p}{k_0} \right),
 \label{eq:sigmascale}
 \end{eqnarray}

where 

\begin{equation}
\nu_0 = \frac{m k_0}{2 \pi}, \qquad \epsilon_0 = \frac{k_0^2}{2m}.
\end{equation}
In  Appendix~B we give explicit expressions for the dimensionless scaling
functions $\tilde{\nu}_{\lambda} ( \tilde{\epsilon}, \tilde{p} )$ and
 $\tilde{\sigma}_{\lambda} ( \tilde{\epsilon}, \tilde{p} )$ as one-dimensional integrals.
In fact, for negative $\tilde{\epsilon}$ the remaining angular
integration in  the expressions for $\tilde{\nu}_{-} ( \tilde{\epsilon}, \tilde{p} )$ and
 $\tilde{\sigma}_{-} ( \tilde{\epsilon}, \tilde{p} )$ can also be done analytically,
see Eqs.~(\ref{eq:nunegativexact}) and (\ref{eq:sigmanegativexact}).
Graphs of the scaling functions are shown in Fig.~\ref{fig:3}.
The behavior of the spectral densities $\nu_- ( \epsilon , p )$ and
$\sigma_- ( \epsilon , p )$ associated with the negative energy branch is
rather interesting. Both functions vanish if $\epsilon$ is smaller than the
lower threshold energy
 \begin{equation}
 \epsilon_- = - \epsilon_0 + \Sigma_0 - v_0 p.
 \end{equation}
Recall that in the absence of an external magnetic field
$ p  =    - \Sigma_x / v_0 = (-g_\bot) M / v_0$.
For energies slightly above the lower threshold we obtain from
Eqs.~(\ref{eq:tildenuasym}, \ref{eq:tildesigmaasym}),
 \begin{eqnarray}
 \frac{\nu_- ( \epsilon , p )}{\nu_0} & \sim &  \frac{\sqrt{ 2 ( 1 + \tilde{p} ) }}{\pi}
\sqrt{ \frac{  \epsilon - \epsilon_{-}}{ v_0 p } },
 \label{eq:numias}
 \\
 \frac{\sigma_- ( \epsilon , p )}{\nu_0} & \sim &  \frac{\sqrt{  1 + \tilde{p}  }}{3 \pi}
\left[ \frac{  \epsilon - \epsilon_{-}}{ v_0 p }  \right]^{3/2},
 \label{eq:sigmamias}
 \end{eqnarray}
where $\tilde{p} = p / k_0$.
Eqs.~(\ref{eq:numias}, \ref{eq:sigmamias}) can also be derived by expanding the energy dispersion
$E_{\bd{k} , - }$ of the lower branch
around the minimum
$\bd{k}_{0} =  k_0 \hat{\bd{x}}$ to quadratic order,
 \begin{equation}
 E_{\bd{k}_{0} + \bd{q}, - } \approx \epsilon_{-} + \frac{ q_x^2 + q_z^2}{2m} + \frac{p}{k_0 + p}
 \frac{ q_y^2}{2m} .
 \label{eq:anisodisp}
 \end{equation}
This approximation is only accurate for
 $|E_{\bd{k}_{0} + \bd{q}, - } - \epsilon_{-} | \ll 2 v_0 p $;
the energy surface $E_{\bd{k} , - } = \epsilon$ can then be approximated by an ellipsoid.
However, for higher energies the topology of the energy surface changes, as
illustrated in Fig.~\ref{fig:energysurface}.
With increasing energy the ellipsoid distorts into a
bean-shaped surface until the two ends of the bean meet at a critical energy.
For higher energies, a hole emerges in the energy surface so that it assumes the
topology of a torus. At the critical energy $\epsilon_{\ast}$ where the topology
of the energy surface changes the spectral densities have a cusp. 
From the exact expressions
for the spectral densities given in Appendix~B
it is easy to see that the critical energy is 
 \begin{equation}
 \epsilon_{\ast} = - \epsilon_0 + \Sigma_0 + v_0 p =
 \epsilon_- + 2 v_0 p.
 \end{equation}
A similar transition in the topology of the Fermi surface of metals
as a function of external pressure has been discussed a long time ago by
Lifshitz \cite{Lifshitz60}.  Close to such a transition, Lifshitz predicted anomalies
in the thermodynamics and the kinetics of the  electrons. 
Therefore we expect that in phases with
spontaneous transverse ferromagnetism
the kinetics of spin-orbit
coupled bosons with energies close
to $\epsilon_{\ast}$ is rather unusual.

\subsection{Solution of the mean-field equations}

For the numerical solution of the above mean-field equations
it is useful to introduce the dimensionless density,
magnetization, susceptibility, and interaction as follows,
 \begin{subequations}
 \begin{eqnarray}
   \tilde{\rho} & = & \frac{\rho}{\nu_0 \epsilon_0 } = \frac{4 \pi \rho }{ k_0^3},
 \\
 \tilde{M}  & = &  \frac{M}{\nu_0 \epsilon_0 } = \frac{4 \pi  M }{ k_0^3},
 \\
 \tilde{\chi}_{\bot} & = & \frac{ \chi_{\bot} }{\nu_0},
 \\
 \tilde{g}_{\bot} & = & \nu_0 g_{\bot}.
 \end{eqnarray}
 \end{subequations}
We also introduce the dimensionless energy 
$\omega = ( \epsilon - \epsilon_- )/ \epsilon_0$
which is measured
relative to the bottom of the lower helicity branch, and define
 \begin{eqnarray}
 \bar{\nu}_{\lambda} ( {\omega}, \tilde{p} ) 
 & = & \frac{ \nu_\lambda (  \epsilon_- +
 \epsilon_0  {\omega} ,  k_0 \tilde{p} ) }{\nu_0}
 =  \tilde{\nu}_\lambda \left( - ( 1 + \tilde{p})^2 + {\omega}  , 
 \tilde{p} \right) ,
 \nonumber
 \\
 & &
 \\
 \bar{\sigma}_{\lambda} ( {\omega}, \tilde{p} ) 
 & = & \frac{ \sigma_\lambda (  \epsilon_- +
 \epsilon_0  {\omega} ,  k_0 \tilde{p} ) }{\nu_0}
 =  \tilde{\sigma}_\lambda \left( - ( 1 + \tilde{p})^2 + {\omega}  , 
 \tilde{p} \right) .
 \nonumber
 \\
 & &
 \end{eqnarray}
Finally, we introduce the dimensionless temperature 
 \begin{eqnarray}
 \tau & = & T / \epsilon_0,
 \end{eqnarray}
and the fugacity
 \begin{equation}
 z  =  e^{  ( \mu - \epsilon_- )/T}.
 \end{equation}
With this notation the density equation (\ref{eq:rhoequation}) can be written as
 \begin{equation}
 \tilde{\rho} = \int_0^{\infty} d \omega 
 \sum_{\lambda}
 \bar{\nu}_{\lambda} ( \omega , \tilde{p} ) \frac{ z}{e^{  \omega / \tau  } - z },
 \label{eq:rhotildedef}
 \end{equation}
while the self-consistency equation (\ref{eq:selfchi}) for the
dimensionless order parameter $\tilde{p} = p / k_0 = - \tilde{g}_{\bot}
 \tilde{M}/2$ becomes
\begin{equation}
  \tilde{\chi}_{\bot} ( \tau, z , \tilde{p} )  =-  1 /   \tilde{g}_{\bot} ,
 \label{eq:selfcon2}
 \end{equation}
with
 \begin{equation}
\tilde{\chi}_{\bot} = -  \frac{1}{2 \tau}
 \int_0^{\infty}  d \omega
 \sum_{\lambda} \lambda
 \bar{\sigma}_{\lambda} ( \omega , \tilde{p} ) \frac{ z  e^{\omega / \tau } 
 }{[ e^{  \omega / \tau  } - z ]^2 }.
 \label{eq:tildechi}
 \end{equation}
Finally, the  dimensionless free energy $f ( \tau , \tilde{\rho} , \tilde{p} ) = 
(\Omega + \mu N ) / ( V \nu_0 \epsilon_0^2 )$ can be written as
 \begin{eqnarray}
 f ( \tau , \tilde{\rho} , \tilde{p} ) & = & \tau \int_0^{\infty} d \omega 
 \sum_{\lambda} \bar{\nu}_{\lambda} ( \omega , \tilde{p} )
 \ln \left[ 1 - z e^{ - \omega / \tau } \right] 
 \nonumber
 \\
 & - &  \frac{ \tilde{g}_{\parallel}}{2}  \tilde{\rho}^2 - \tilde{g}_{\bot} \left( 
 \frac{\tilde{\rho}^2}{4} + \tilde{M}^2 \right)  + \tilde{\mu} \tilde{\rho},
 \label{eq:freedef}
 \end{eqnarray}
where $\tilde{\mu} = \mu / \epsilon_0$.
Note that at constant density we should determine $\mu$ as a function
of $\tilde{\rho}$  and $\tilde{p}$.

\begin{figure}[tb]    
  \centering
  \subfigure[]{\includegraphics[width=1 \linewidth]{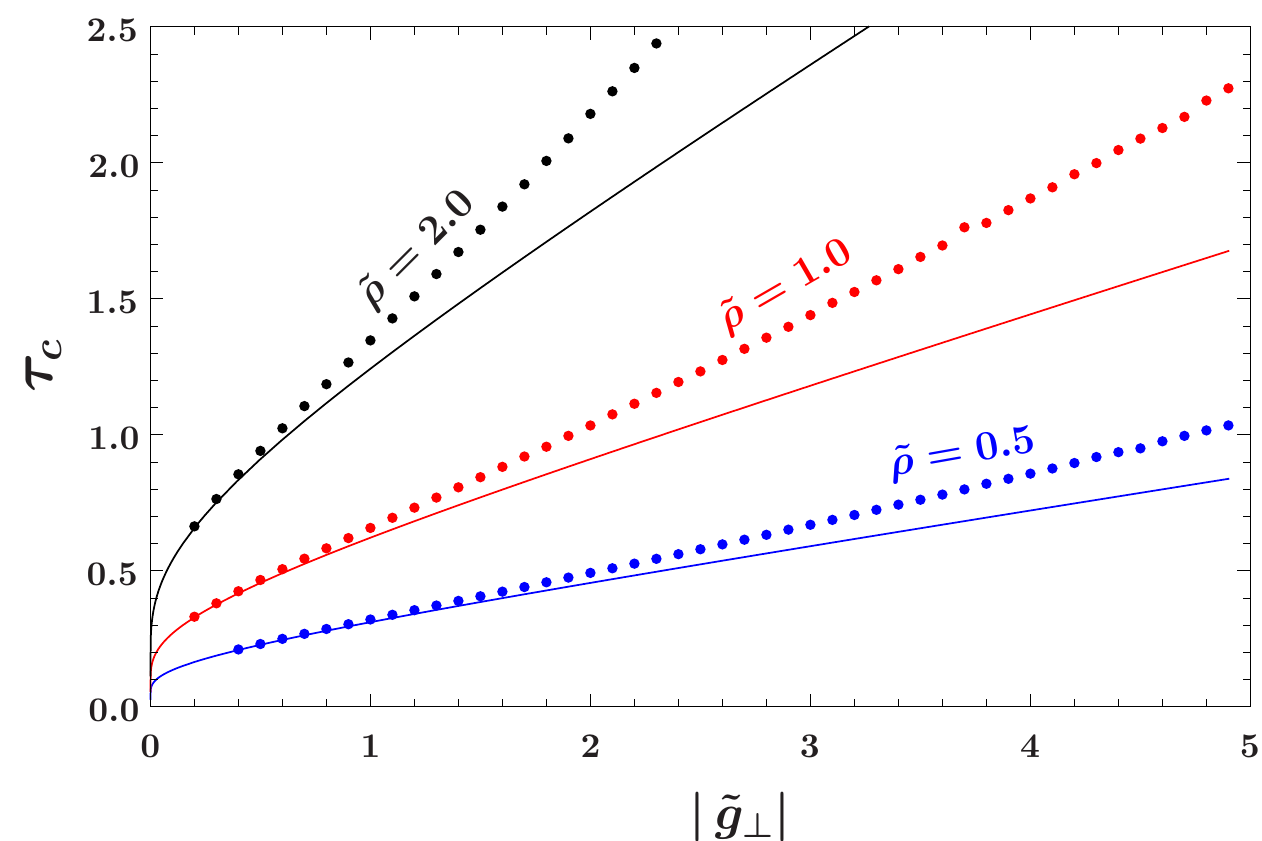}}
  \subfigure[]{\includegraphics[width=1 \linewidth]{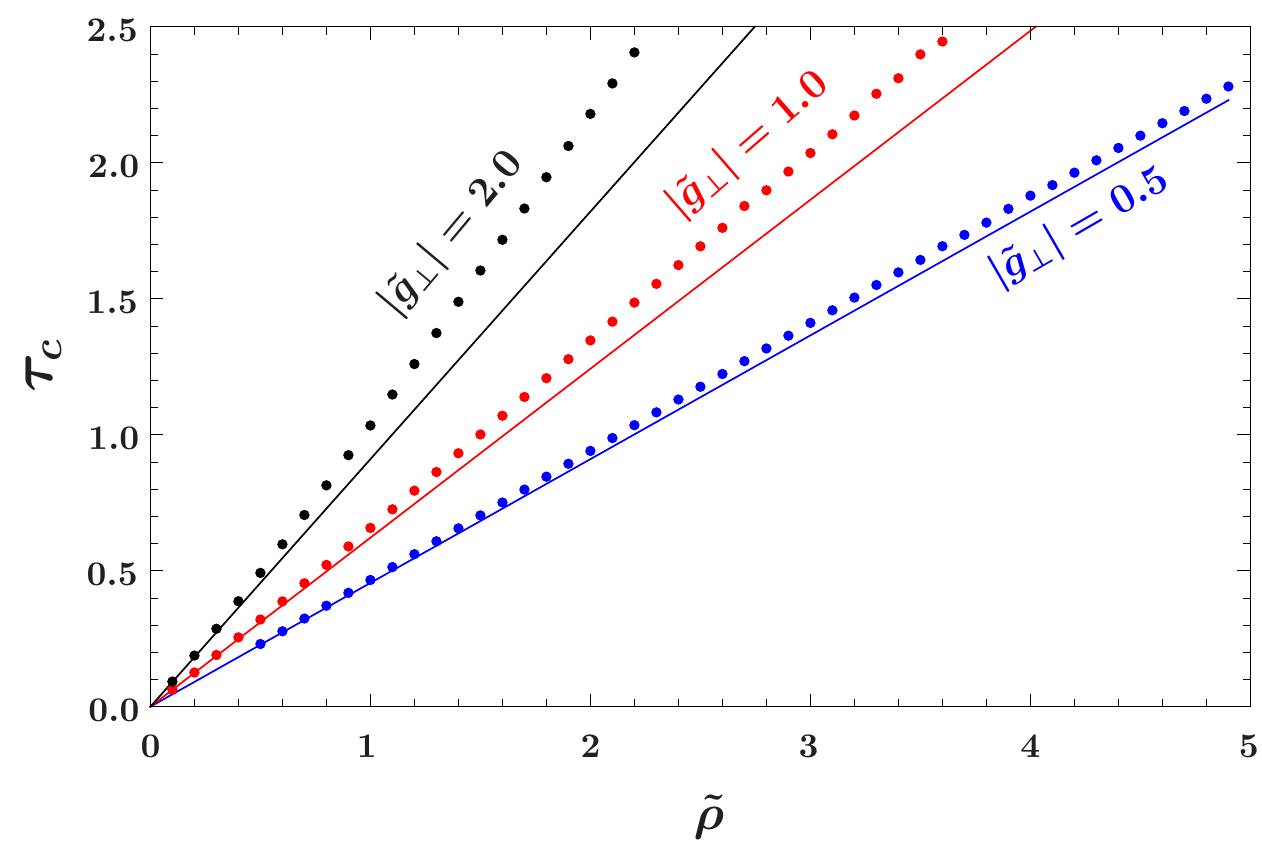}}
  \caption{\label{fig:5}(Color online) %
(a) Critical temperature for transverse ferromagnetism 
as a function of the dimensionless coupling constant $ | \tilde{g}_{\bot} |$
for three different densities. The dots have been obtained from the numerical
solution of the mean-field equations (\ref{eq:chicrit}) without further approximation,
while the solid lines represent the low-temperature 
approximation (\ref{eq:tauc}).
~(b) Critical temperature as a function of density for different values
of the interaction. 
} 
 \label{fig:tauc}
\end{figure}

Let us first discuss the critical temperature $\tau_c = T_c / \epsilon_0$ below which the system exhibits
spontaneous transverse ferromagnetism. 
According to Eq.~(\ref{eq:selfcon2}), for a given density $\tilde{\rho}$ the critical temperature
is determined by
 \begin{equation}
 \tilde{\chi}_{\bot} ( \tau_c , z_c ( \tau_c, \tilde{\rho}  ), \tilde{p} =0 ) = - 1 / \tilde{g}_{\bot},
 \label{eq:chicrit}
 \end{equation}
where the  fugacity $z_c ( \tau_c , \tilde{\rho})$ at the critical point is determined by
 \begin{equation}
\tilde{\rho} = \int_0^{\infty} d \omega 
 \sum_{\lambda}
 \bar{\nu}_{\lambda} ( \omega, 0 ) \frac{ z_c}{e^{  \omega / \tau_c  } - z_c }.
 \label{eq:fugcrit}
 \end{equation}
Numerical results for $\tau_c$ as a function of  $|\tilde{g}_{\bot} |$
for different densities are shown in Fig.~\ref{fig:tauc} (a), while
in Fig.~\ref{fig:tauc} (b) we show the critical temperature as a function
of density for different values of $| \tilde{g}_{\bot}|$. The numerical results in general are obtained without
approximating the spectral densities and choosing $g_\|$ such that $\Sigma_0$ vanishes.   
Note that for small $ | \tilde{g}_{\bot}|$ the critical temperature approaches
zero with infinite slope.  In this regime it is easy to obtain an analytic expression
for the critical temperature. 
Assuming that the temperature is small compared with $\epsilon_0$ 
(corresponding to $\tau \ll 1$)
we may neglect the contribution
of the upper helicity branch and
approximate the spectral densities
by their leading asymptotics for frequencies close to the bottom of the lower helicity branch
given in Eqs.~(\ref{eq:nu0}) and (\ref{eq:sigma0}).
In the symmetric phase where $ \tilde{p} =0$ the density equation  (\ref{eq:rhotildedef})
then reduces to
 \begin{equation}
 \tilde{\rho} =  \int_0^{\infty} d \omega \frac{ z}{ e^{\omega / \tau } - z }
 = - \tau \ln ( 1 - z ).
 \end{equation} 
We conclude that in the regime where the critical temperature $\tau_c$
is small compared with unity, the critical fugacity is given by
 \begin{equation}
  z_c = 1- e^{- \tilde{\rho} / \tau_c }.
 \end{equation}
For $\tau_c \ll \tilde{\rho}$ this is exponentially close to unity,
which is a consequence
of the finite density of states of the spin-orbit coupled Bose gas close to the 
bottom of the lower energy branch.
To determine the critical temperature as a function of the density,
we calculate the transverse susceptibility from
Eq.~(\ref{eq:tildechi}) with $\tilde{p} =0$, using the
approximation (\ref{eq:sigma0}) for the spectral density 
$   \bar{\sigma}_- ( \omega , 0 ) =  \tilde{\sigma}_- ( -1 + \omega , \tilde{p} =0)$
for frequencies close to the bottom of the lower helicity branch,
 \begin{eqnarray}
 \tilde{\chi}_{\bot} & = & \frac{1}{4}
 \int_0^{\infty} d \omega \frac{ z e^{\omega / \tau }   }{ [e^{\omega / \tau } - z]^2 }
 = \frac{1}{4} \frac{ z}{1-z} = \frac{1}{4} \left[ e^{ \tilde{\rho} / \tau } -1 \right].
 \nonumber
 \\
 & &  
 \label{eq:chilow}
 \end{eqnarray}
Hence, in the paramagnetic phase the transverse susceptibility 
becomes exponentially large at low temperatures. As a consequence, 
for any finite attractive interaction $ g_{\bot} < 0$,
we can find a solution of the self-consistency equation  (\ref{eq:selfcon2})
at sufficiently low temperatures.   Combining Eqs.~(\ref{eq:chicrit}) and (\ref{eq:chilow})
we obtain for the critical temperature
 \begin{equation}
\tau_c = \frac{\tilde{\rho}}{ \ln \left( 1 + 4 / | \tilde{g}_{\bot} | \right)}.
 \label{eq:tauc}
 \end{equation}
From the derivation of this expression it is clear that
Eq.~(\ref{eq:tauc})  is valid as long as $\tau_c \ll 1$, 
which can always be satisfied for sufficiently small densities.
The approximation (\ref{eq:tauc}) corresponds to the  solid
lines in Fig.~\ref{fig:tauc}.

\begin{figure}[tb]    
  \centering
  \includegraphics[keepaspectratio,width=\linewidth]{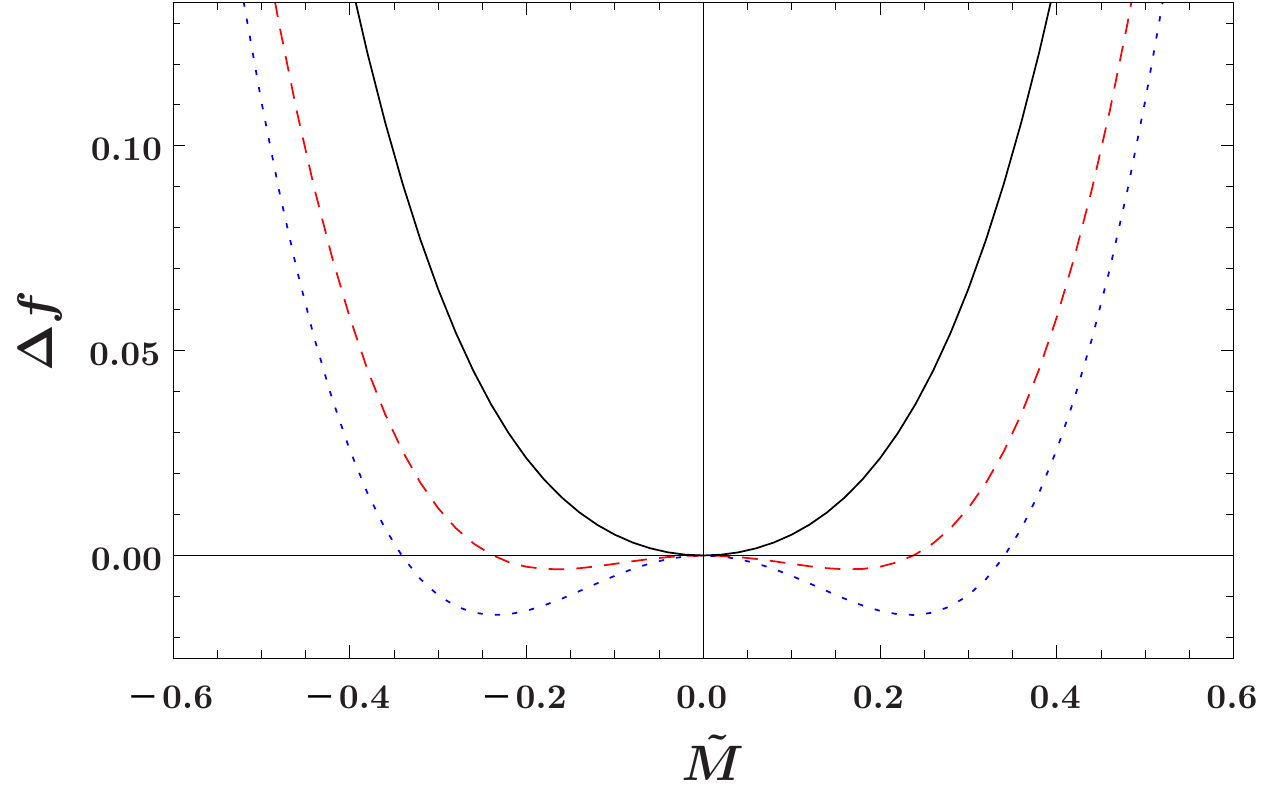}
  \caption{\label{fig:6}(Color online) %
 Dimensionless free energy $\Delta f$, defined in equation (\ref{eq:freeenergy}), as a function of 
the dimensionless magnetization $\tilde{M}$ for three different temperatures.
$\tau/\tau_c=0.9$ (dotted line);  $\tau/\tau_c=0.95$ (dashed line); $\tau/\tau_c=1.1$ (solid line). 
For all plots  we have used $|\tilde{g}_\bot|=5$ and $\tilde{\rho}=1$. 
 \label{fig:freeenergy}
 } 
\end{figure}

\begin{figure}[tb]    
  \centering
  \includegraphics[keepaspectratio,width=\linewidth]{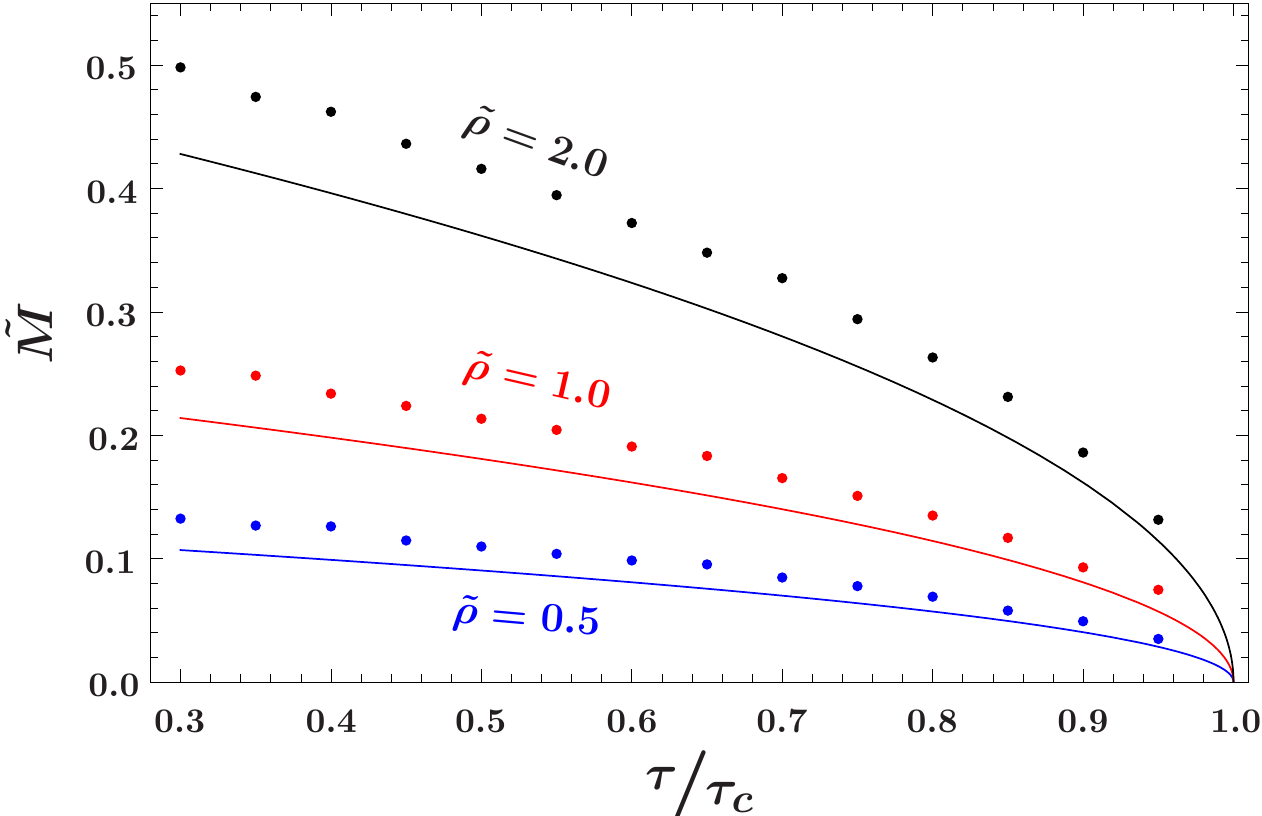}
  \caption{\label{fig:magnetisationplot}(Color online) %
Transverse magnetization as a function of temperature. The plots are for $|\tilde{g}_\bot|=1$. The dots are obtained numerically,
 while the solid lines correspond to the analytic result (\ref{eq:Mcrit}) which is only valid for $\tau_c-\tau\ll \tau_c \ll 1$. 
} 
 \label{fig:magnetization}
\end{figure}

Next, let us discuss the low-temperature phase $\tau < \tau_c$ with spontaneous
transverse ferromagnetism.
In Fig.~\ref{fig:freeenergy}
we show the change 
\begin{equation}
\Delta f = f ( \tau , \tilde{\rho} , \tilde{p} ) - f ( \tau , \tilde{\rho} , 0 )
\label{eq:freeenergy}
\end{equation}
in the dimensionless free energy defined in Eq.~(\ref{eq:freedef})
as a function of the order parameter $\tilde{M}  = 2 \tilde{p} / | \tilde{g}_{\bot} |$
for three different temperatures.
For $ \tau < \tau_c$ the free energy continuously develops  two degenerate
minima, corresponding to the
Hartree-Fock solutions $ \pm \tilde{M}$.
The phase transition to the magnetic state is therefore second order.
The transverse magnetization as a function of temperature for three different densities, obtained numerically, is shown in Fig.~\ref{fig:magnetization}.
In the regime where $\tau_c \ll 1$ 
the behavior of the magnetization for temperatures close to the critical
temperature can be calculated analytically, as shown in Appendix C.
We find
\begin{equation}
 \tilde{M} \sim  \frac{\tilde{\rho}}{     \sqrt{\left( 8 + \frac{3}{2} | \tilde{g}_{\bot} | 
 \right) \ln [ 1 + \frac{4}{ | \tilde{g}_{\bot} | } ] }} 
 \sqrt{ \frac{ \tau_c - \tau }{\tau_c}}.
 \label{eq:Mcrit}
 \end{equation}
The reason we obtain the usual mean-field exponent $\beta =1/2$ is of course
related to the fact that our calculation is based on the Hartree-Fock approximation.

\section{Summary and conclusions}

In summary, we have shown that in the spinor Bose gas
with Rashba-type spin-orbit coupling an arbitrarily weak attractive
interaction between bosons with opposite spin triggers
a ferromagnetic instability for temperatures below some finite temperature $T_c$ if the density
of the bosons is fixed.
Note that spontaneous ferromagnetism in electronic systems
usually appears only if the relevant interaction exceeds a finite
threshold \cite{Moriya85}. 
The fact that in the spinor Bose gas 
such a threshold does not exist is related to
the singularity of the Bose function for 
small energies in combination with the finite density of states
in the non-magnetic phase
due to the spin-orbit coupling of the Rashba-type.
Because for  $ T < T_c$, the density of states exhibits the usual $\sqrt{\omega}$-behavior
at low energies, at some temperature below $T_c$
the bosons eventually condense into the single-particle state
with the lowest momentum $\bd{k}_{0}$, which is unique
in the ferromagnetic phase.
In the weak coupling regime, we may estimate the 
critical temperature for BEC by using the critical temperature
for free bosons with anisotropic dispersion given by Eq.~(\ref{eq:anisodisp}),
 \begin{eqnarray}
 T_{\rm BEC} & = & \frac{ 2 \pi }{m} 
 \left( \frac{ \rho}{\zeta ( 3/2) } \right)^{2/3} \left(\frac{ p}{k_0 + p }  \right)^{1/3}
 \nonumber
 \\
 & = & \frac{ 2 \pi }{m} 
 \left( \frac{ \rho}{\zeta ( 3/2) } \right)^{2/3}
 \left( \frac{ | g_{\bot} | M }{k_0^2/m    +   | g_{\bot} | M   } \right)^{1/3}. 
 \hspace{10mm}
 \end{eqnarray}
For consistency, we should require that $T_{\rm BEC} < T_c$, 
which is satisfied at weak coupling where $p \ll k_0$.
We thus conclude that the ground state of the spinor Bose gas with
Rashba spin-orbit coupling and attractive interaction $g_{\bot} < 0$
is a ferromagnetic Bose-Einstein condensate.

In principle, an attractive interaction $g_{\bot} < 0$ could also 
trigger an instability in the particle-particle channel, 
leading to a pair condensate at low temperatures. However, as shown 
in Appendix D, at least for sufficiently low densities
the ferromagnetic instability has a higher critical temperature
if the relevant coupling constants have the same order of magnitude.

Let us also point out that the constant $g_{\bot}$
in our mean-field calculation should be considered as an effective
low-energy interaction in the opposite-spin
particle-hole channel. This interaction 
includes renormalization effects from high-energy fluctuations
in all channels, so that it is in principle possible that the
effective low-energy coupling is attractive even if we start
from a repulsive bare interaction. Note that recently
Gopalakrishnan {\it{et al.}} \cite{Gopalakrishnan11} performed
a momentum shell renormalization group calculation
for the spin-orbit coupled Bose gas
whose dispersion exhibits a minimum on a circle
in momentum space; at finite temperature they found evidence for an attractive renormalized coupling in the particle-particle channel, although
the corresponding bare interaction is repulsive, suggesting an instability towards pair condensation.
However, 
a proper renormalization group
calculation of the effective low-energy coupling 
of the spinor Bose gas with Rashba-type spin-orbit coupling, taking
high-energy fluctuations in all channels 
and all relevant and marginal couplings
consistently into account, still
remains to be done. Given the fact that
the system belongs to the Brazovskii universality class,
the rescaling step in the renormalization group
transformation is non-trivial and all two-body scattering processes
describing particles with momenta on the
low-energy manifold are marginal and should be retained.
It  should be advantageous
to use the well developed machinery of the functional renormalization
group \cite{Kopietz10,Kopietz01,Metzner12} to carry out such a calculation.

\section*{ACKNOWLEDGMENTS}

This work was financially supported by
SFB/TRR49. P. Z. would like to thank M. Zentkov\'a 
and A. Lencsesov\'a for financial support from project
EDUFYCE (ITMS code 26110230034).

\begin{appendix}

\renewcommand{\theequation}{A\arabic{equation}}

\section*{APPENDIX A: Interactions in the helicity basis}
\setcounter{equation}{0}

In this appendix we explicitly give the interaction part 
of the boson Hamiltonian (\ref{eq:H1}) 
in the helicity basis. For simplicity, we assume
momentum independent bare
interactions, see Eq.~(\ref{eq:gdef}).
To transform the interaction from the spin basis to the helicity basis,
it is useful to introduce the notation
$a_{\bd{k}} = a_{\bd{k} -}$ and $b_{\bd{k}} = e^{ - i \varphi_{\bd{k}} } a_{\bd{k} +}$,
so that our transformation (\ref{eq:trafo}) to the helicity basis can be written as
 \begin{eqnarray}
 a_{\bd{k} \uparrow } & = & \frac{1}{\sqrt{2}} \left[ a_{\bd{k} -} -  e^{ - i    \varphi_{\bd{k}}} 
a_{\bd{k} +  }
 \right] =
\frac{1}{\sqrt{2}} \left[ a_{\bd{k} } -  b_{\bd{k}  }
 \right],
 \\
a_{\bd{k} \downarrow } & = &
\frac{ 1}{\sqrt{2}} 
 \left[  e^{ i \varphi_{\bd{k}} }   a_{\bd{k} -} +  a_{\bd{k} + } \right] =
 \frac{  e^{ i \varphi_{\bd{k}} }}{\sqrt{2}} 
 \left[    a_{\bd{k} } +  b_{\bd{k}} \right].
 \end{eqnarray}
Defining the coupling constants $g_{\uparrow}$, $g_{\downarrow}$ and $g_{\bot}$
as in Eq.~(\ref{eq:gdef}), we find that in the helicity basis  the interaction part (\ref{eq:Hint}) 
of our Hamiltonian can be written as
 \begin{widetext}
 \begin{eqnarray}
 {\cal{H}}_{\rm int}
& = &\frac{1}{ V} \sum_{ 
 \bd{k}_1^{\prime} \bd{k}_2^{\prime} \bd{k}_{2} \bd{k}_1 }
 \delta_{ \bd{k}_1^{\prime}+ \bd{k}_2^{\prime}, \bd{k}_{2}+ \bd{k}_1}
 \Bigl\{  \frac{1}{(2!)^2} 
 \Bigl[ \Gamma_0^{\bar{a} \bar{a} aa} 
 ( 1^{\prime} ,  2^{\prime};  {2},  1 ) 
  a^{\dagger}_{ 1^{\prime}} a^{\dagger}_{ 2^{\prime}} a_{2} a_1
+ \Gamma_0^{\bar{b} \bar{b} bb} 
 ( 1^{\prime} ,  2^{\prime};  {2},  1 ) 
  b^{\dagger}_{ 1^{\prime}} b^{\dagger}_{ 2^{\prime}} b_{2} b_1
 \nonumber
 \\
 & & \hspace{46mm} + \Gamma_0^{\bar{a} \bar{a} bb} 
 ( 1^{\prime} ,  2^{\prime};  {2},  1 ) 
  a^{\dagger}_{ 1^{\prime}} a^{\dagger}_{ 2^{\prime}} b_{2} b_1
+ \Gamma_0^{\bar{b} \bar{b} aa} 
 ( 1^{\prime} ,  2^{\prime};  {2},  1 ) 
  b^{\dagger}_{ 1^{\prime}} b^{\dagger}_{ 2^{\prime}} a_{2} a_1
 \Bigr]
 \nonumber
 \\
 &  & \hspace{40mm} + 
 \frac{1}{2!} 
 \Bigl[ \Gamma_0^{\bar{a} \bar{a} ab} 
 ( 1^{\prime} ,  2^{\prime};  {2} ,  1 ) 
  a^{\dagger}_{ 1^{\prime}} a^{\dagger}_{ 2^{\prime}} a_{2} b_1
+ \Gamma_0^{\bar{b} \bar{b} ba} 
 ( 1^{\prime} ,  2^{\prime};  {2} ,  1 ) 
  b^{\dagger}_{ 1^{\prime}} b^{\dagger}_{ 2^{\prime}} b_{2} a_1
 \nonumber
 \\
 & & \hspace{46mm} + \Gamma_0^{\bar{b} \bar{a} aa} 
 ( 1^{\prime} ,  2^{\prime};  {2},  1 ) 
  b^{\dagger}_{ 1^{\prime}} a^{\dagger}_{ 2^{\prime}} a_{2} a_1
+ \Gamma_0^{\bar{a} \bar{b} bb} 
 ( 1^{\prime} ,  2^{\prime};  {2},  1 ) 
  a^{\dagger}_{ 1^{\prime}} b^{\dagger}_{ 2^{\prime}} b_{2} b_1
 \Bigr]
 \nonumber
 \\
 &  &  \hspace{40mm} + 
   \Gamma_0^{\bar{a} \bar{b} ba} 
 ( 1^{\prime} ,  2^{\prime};  {2} ,  1 ) 
  a^{\dagger}_{ 1^{\prime}} b^{\dagger}_{ 2^{\prime}} b_{2} a_1
 \Bigr\}
 \nonumber
 \\
 & = &\frac{1}{ V} \sum_{ 
 \bd{k}_1^{\prime} \bd{k}_2^{\prime} \bd{k}_{2} \bd{k}_1 }
 \delta_{ \bd{k}_1^{\prime}+ \bd{k}_2^{\prime}, \bd{k}_{2}+ \bd{k}_1}
 \Bigl\{  \frac{1}{(2!)^2} 
 U_1 ( \bd{k}_1^{\prime} ,  \bd{k}_2^{\prime};  \bd{k}_{2},  \bd{k}_1 )
  [ a^{\dagger}_{ 1^{\prime}} a^{\dagger}_{ 2^{\prime}} a_{2} a_1 +
 b^{\dagger}_{ 1^{\prime}} b^{\dagger}_{ 2^{\prime}} b_{2} b_1  ]
 \nonumber
 \\
 & & 
\hspace{41mm} 
 + \frac{1}{(2!)^2} U_2 ( \bd{k}_1^{\prime} ,  \bd{k}_2^{\prime};  \bd{k}_{2},  \bd{k}_1 )
  [ a^{\dagger}_{ 1^{\prime}} a^{\dagger}_{ 2^{\prime}} b_{2} b_1 +
 b^{\dagger}_{ 1^{\prime}} b^{\dagger}_{ 2^{\prime}} a_{2} a_1  ]
\nonumber
 \\
 & & 
\hspace{41mm} 
 + \frac{1}{2!} U_3 ( \bd{k}_1^{\prime} ,  \bd{k}_2^{\prime};  \bd{k}_{2} ;  \bd{k}_1 )
  [ a^{\dagger}_{ 1^{\prime}} a^{\dagger}_{ 2^{\prime}} a_{2} b_1 +
 b^{\dagger}_{ 1^{\prime}} b^{\dagger}_{ 2^{\prime}} b_{2} a_1  ]
 \nonumber
 \\
 & & 
\hspace{41mm} 
+ \frac{1}{2!} U_4 ( \bd{k}_1^{\prime} ;  \bd{k}_2^{\prime};  \bd{k}_{2} ,  \bd{k}_1 )
  [ b^{\dagger}_{ 1^{\prime}} a^{\dagger}_{ 2^{\prime}} a_{2} a_1 +
 a^{\dagger}_{ 1^{\prime}} b^{\dagger}_{ 2^{\prime}} b_{2} b_1  ]
\nonumber
 \\
 & &
\hspace{41mm} 
 +  U_5 ( \bd{k}_1^{\prime} ;  \bd{k}_2^{\prime};  \bd{k}_{2} ;  \bd{k}_1 )
   a^{\dagger}_{ 1^{\prime}} b^{\dagger}_{ 2^{\prime}} b_{2} a_1 
 \Bigr\}. \hspace{7mm}
 \label{eq:Hintlong}
\end{eqnarray}
The properly symmetrized interaction vertices are 
\begin{subequations}
 \begin{eqnarray}
  \Gamma_0^{\bar{a} \bar{a} aa} 
 ( 1^{\prime} ,  2^{\prime};  {2},  1 )    =      \Gamma_0^{ \bar{b} \bar{b} bb} 
 ( 1^{\prime} ,  2^{\prime};  {2},  1 )   
 & \equiv & U_1 ( \bd{k}_1^{\prime} ,  \bd{k}_2^{\prime};  \bd{k}_{2},  \bd{k}_1 ) 
 \nonumber
 \\  
 & = & 
  \frac{g_{\uparrow}}{2} + \frac{g_{\downarrow}}{2} e^{ - i (  \varphi_{1^{\prime}} +
 \varphi_{2^{\prime}} - \varphi_2 - \varphi_1 )}
 +  \frac{g_{\bot}}{4} ( e^{ - i \varphi_{1^{\prime}} } + e^{ - i \varphi_{2^{\prime}}} )
  ( e^{ i \varphi_{1} } + e^{ i \varphi_{2}} ),
 \\
 \Gamma_0^{\bar{a} \bar{a} bb} 
 ( 1^{\prime} ,  2^{\prime};  {2},  1 )   =     \Gamma_0^{ \bar{b} \bar{b} aa} 
 ( 1^{\prime} ,  2^{\prime};  {2},  1 )   
& \equiv &  U_2 ( \bd{k}_1^{\prime} ,  \bd{k}_2^{\prime};  \bd{k}_{2},  \bd{k}_1 ) 
 \nonumber
 \\
& = &
  \frac{g_{\uparrow}}{2} + \frac{g_{\downarrow}}{2} e^{-  i (  \varphi_{1^{\prime}} +
 \varphi_{2^{\prime}} - \varphi_2 - \varphi_1 )}
  - \frac{g_{\bot}}{4} ( e^{ - i \varphi_{1^{\prime}} } + e^{ - i \varphi_{2^{\prime}}} )
  ( e^{ i \varphi_{1} } + e^{ i \varphi_{2}} ),
 \hspace{7mm}
 \\
\Gamma_0^{\bar{a} \bar{a} ab} 
 ( 1^{\prime} ,  2^{\prime};  {2} ,  1 ) 
 =
  \Gamma_0^{\bar{b} \bar{b} ba} 
 ( 1^{\prime} ,  2^{\prime};  {2} ,  1 ) 
& \equiv & U_3 ( \bd{k}_1^{\prime} ,  \bd{k}_2^{\prime};  \bd{k}_{2} ;  \bd{k}_1 ) 
 \nonumber
 \\
 & = &
  - \frac{g_{\uparrow}}{2} + \frac{g_{\downarrow}}{2} e^{ - i (  \varphi_{1^{\prime}} +
 \varphi_{2^{\prime}} - \varphi_2 - \varphi_1 )}
 + 
  \frac{g_{\bot}}{4} ( e^{ - i \varphi_{1^{\prime}} } + e^{ - i \varphi_{2^{\prime}}} )
  ( e^{ i \varphi_{1} } - e^{ i \varphi_{2}} ),
 \hspace{7mm}
 \\
\Gamma_0^{\bar{b} \bar{a} aa} 
 ( 1^{\prime} ,  2^{\prime};  {2},  1 ) 
=  \Gamma_0^{\bar{a} \bar{b} bb} 
 ( 1^{\prime} ,  2^{\prime};  {2},  1 ) 
& \equiv & U_4 ( \bd{k}_1^{\prime} ;  \bd{k}_2^{\prime};  \bd{k}_{2} ,  \bd{k}_1 ) 
 \nonumber
 \\
 & = & 
  - \frac{g_{\uparrow}}{2} + \frac{g_{\downarrow}}{2} e^{ - i (  \varphi_{1^{\prime}} +
 \varphi_{2^{\prime}} - \varphi_2 - \varphi_1 )}
  +
  \frac{g_{\bot}}{4} ( e^{ - i \varphi_{1^{\prime}} } - e^{ - i \varphi_{2^{\prime}}} )
  ( e^{ i \varphi_{1} } + e^{ i \varphi_{2}} ),
 \hspace{7mm}
 \\
  \Gamma_0^{\bar{a} \bar{b} ba} 
 ( 1^{\prime} ,  2^{\prime};  {2} ,  1 ) 
 & \equiv & U_5 ( \bd{k}_1^{\prime} ;  \bd{k}_2^{\prime};  \bd{k}_{2} ;  \bd{k}_1 ) 
 \nonumber
 \\
 & = &
  \frac{g_{\uparrow}}{2} + \frac{g_{\downarrow}}{2} e^{ - i (  \varphi_{1^{\prime}} +
 \varphi_{2^{\prime}} - \varphi_2 - \varphi_1 )}
  +
  \frac{g_{\bot}}{4} ( e^{ - i \varphi_{1^{\prime}} } - e^{ - i \varphi_{2^{\prime}}} )
  ( e^{ i \varphi_{1} } - e^{ i \varphi_{2}} ).
 \hspace{7mm}
 \end{eqnarray}
 \end{subequations}
\end{widetext}
The parametrization of the vertices  in terms of five functions $U_1, \ldots , U_5$ 
is similar to the parametrization used by Ozawa and Baym \cite{Ozawa11}; however, we find it
convenient to introduce slightly different numerical prefactors
in the second line of Eq.~(\ref{eq:Hintlong})
in order to simplify the combinatorial factors due the permutation symmetries 
of the vertices in higher order calculations.
Note that the interaction vertices are invariant under arbitrary rotations
around the $z$ axis, corresponding to a shift $\varphi_{\bd{k}} \rightarrow
 \varphi_{\bd{k}} + \alpha$ in all angles.

\section*{APPENDIX B: Spectral densities}

\renewcommand{\theequation}{B\arabic{equation}}
\setcounter{equation}{0}

In this appendix we discuss the dimensionless scaling functions
$\tilde{\nu}_{\lambda} ( \tilde{\epsilon} , \tilde{p} )$ and
$\tilde{\sigma}_{\lambda} ( \tilde{\epsilon} , \tilde{p} )$ 
which are defined by writing the density of states and the weighted density of states
in the scaling form (\ref{eq:nuscale}, \ref{eq:sigmascale}).
The scaling function $\tilde{\nu}_{\lambda} ( \tilde{\epsilon} , \tilde{p} )$
for the density of states can be written as
 \begin{widetext}
 \begin{eqnarray}
 \tilde{\nu}_{\lambda} ( \tilde{\epsilon} , \tilde{p} ) & = &  
  \Theta ( \tilde{\epsilon} )
 \left\{ \frac{\sqrt{ \tilde{\epsilon}}}{\pi} +   \int_0^{2 \pi} \frac{ d \varphi}{ 2 \pi}   ( \tilde{p} 
 \cos \varphi - \lambda) 
 \left[ \frac{1}{2} + \frac{1}{\pi} \arctan \left( \frac{  \tilde{p} \cos \varphi - \lambda}{\sqrt{\tilde{\epsilon}}}
 \right)  \right] \right\}
 \nonumber
 \\
 &  + & \Theta ( - \tilde{\epsilon} )  \int_0^{2 \pi} \frac{ d \varphi}{ 2 \pi}
  (  \tilde{p} \cos \varphi - \lambda )
 \Theta (  \tilde{p} \cos \varphi - \lambda - \sqrt{ - \tilde{\epsilon} } ).
 \label{eq:nuscale1int}
 \end{eqnarray}
Recall that in Eqs. (\ref{eq:nuscale}) and (\ref{eq:sigmascale}) the dimensionless variables $\tilde{\epsilon}$ and $\tilde{p}$ represent
 \begin{equation}
\tilde{\epsilon}=\frac{\epsilon-\Sigma_0-\frac{p^2}{2m}}{\epsilon_0}, \qquad \tilde{p}=\frac{p}{k_0}. 
\end{equation}
For negative $\tilde{\epsilon}$ the angular integration in Eq.~(\ref{eq:nuscale1int}) can be done analytically.
For the lower energy branch ($ \lambda = -1$) we obtain
\begin{eqnarray}
 \tilde{\nu}_{-} ( \tilde{\epsilon} , \tilde{p} )
 & = &  \Theta ( 1   - \sqrt{ - \tilde{\epsilon}}  - \tilde{p} )
+ \Theta ( \tilde{p} - | 1 - \sqrt{ - \tilde{\epsilon} }   | )  \frac{1}{\pi}
 \left[ 
 \arccos \left( \frac{ \sqrt{ - \tilde{\epsilon} } -1  }{\tilde{p} } \right)
 +\sqrt{ \tilde{p}^2 -  ( 1 - \sqrt{ - \tilde{\epsilon} }   )^2}
  \right],
 \; \; \; \mbox{for $ \tilde{\epsilon} < 0 $}.
 \label{eq:nunegativexact}
 \end{eqnarray}
The scaling function for the weighted density of states
$\tilde{\sigma}_{\lambda} ( \tilde{\epsilon}, \tilde{p} )$, can be written as
 \begin{eqnarray}
 \tilde{\sigma}_{\lambda} ( \tilde{\epsilon}, \tilde{p} ) & = &  
 \Theta ( \tilde{\epsilon} )
 \int_0^{2 \pi} \frac{ d \varphi}{ 2 \pi}  \frac{\sin^2 \varphi}{2}
 \left\{
 \left[
 3   ( \tilde{p}  \cos \varphi - \lambda )^2 + \tilde{\epsilon} 
 \right]
 \left[ \frac{1}{2} + \frac{1}{\pi} \arctan \left( \frac{  \tilde{p} \cos \varphi - \lambda }{\sqrt{\tilde{\epsilon}}}
 \right)  \right]    + \frac{  3  \sqrt{  \tilde{\epsilon}}   }{\pi}    ( \tilde{p}  \cos \varphi - \lambda )  \right\}
 \nonumber
 \\
 &  + & \Theta ( - \tilde{\epsilon} )   \int_0^{2 \pi} \frac{ d \varphi}{ 2 \pi}
 \frac{\sin^2 \varphi}{2}
  \left[ 3 (  \tilde{p} \cos \varphi - \lambda )^2 + \tilde{\epsilon} \right]
 \Theta (\tilde{p} \cos \varphi - \lambda - \sqrt{ - \tilde{\epsilon} } ).
 \label{eq:sigmascale1int}
 \end{eqnarray}
For negative $\tilde{\epsilon}$ the angular integration can again be performed analytically.
For the lower energy branch ($ \lambda = -1$) we obtain
\begin{eqnarray}
 \tilde{\sigma}_{-} ( \tilde{\epsilon} , \tilde{p} ) & = &  \Theta ( 1   - \sqrt{ - \tilde{\epsilon}}  - \tilde{p} )
 \frac{1}{4} \left( 3 + \tilde{\epsilon} + \frac{3}{4} \tilde{p}^2 \right)
+ \Theta ( \tilde{p} - | 1 - \sqrt{ - \tilde{\epsilon} }   | )  \frac{1}{4 \pi}
 \Biggl[   \left( 3 + \tilde{\epsilon} + \frac{3}{4} \tilde{p}^2 \right)
 \arccos \left( \frac{ \sqrt{ - \tilde{\epsilon} } -1  }{\tilde{p} } \right) 
  \nonumber
 \\
& &
+\sqrt{ \tilde{p}^2 -  ( 1 - \sqrt{ - \tilde{\epsilon} }   )^2}
 \left( \frac{ ( 1 + \tilde{\epsilon} ) ( 1 + \sqrt{ - \tilde{\epsilon} } )}{2 \tilde{p}^2}
 + \frac{13 + 3 \sqrt{ - \tilde{\epsilon} }}{4} \right)
\Biggr],
 \; \; \; \mbox{for $ \tilde{\epsilon} < 0 $}.
 \label{eq:sigmanegativexact}
 \end{eqnarray}
\end{widetext}
Plots of the above scaling functions 
are shown in Fig.~\ref{fig:scalingfuncs}.
The asymptotic behavior slightly above the lower threshold
$ \tilde{\epsilon}_- =  -(1 + \tilde{p})^2$ is,
 \begin{eqnarray}
  \tilde{\nu}_{-} ( \tilde{\epsilon} , \tilde{p} )
  & \sim &  \frac{\sqrt{ 1 + \tilde{p}}}{\pi} \sqrt{ \frac{ \tilde{\epsilon} + ( 1 +
 \tilde{p} )^2}{\tilde{p}}},
 \label{eq:tildenuasym}
 \\
 \tilde{\sigma}_- ( \tilde{\epsilon} , \tilde{p} ) & \sim & \frac{ \sqrt{ 1 + \tilde{p}}    }{3 \pi} 
 \left[ \frac{ \tilde{\epsilon} + ( 1 + \tilde{p} )^2 }{ \tilde{p} } \right]^{3/2}.
 \label{eq:tildesigmaasym}
 \end{eqnarray}
To discuss the thermodynamics close to the critical point, we need the expansion of the density of states and the weighted density of states 
for small $\tilde{p} = p / k_0$ to order $\tilde{p}^2$. We obtain
  \begin{equation}
 \frac{{\nu}_{\lambda} ( \epsilon, p )}{\nu_0}    = \tilde{\nu}_{\lambda}^{(0)} 
 \left( \frac{ \epsilon - \Sigma_0 }{\epsilon_0} \right)
   +   \tilde{p}^2  \tilde{\nu}_{\lambda}^{(2)} 
  \left( \frac{ \epsilon - \Sigma_0 }{\epsilon_0} \right) 
  + {\cal{O}} ( \tilde{p}^4 ),
 \label{eq:nuexpansion}
 \end{equation}
where
 \begin{eqnarray}
 \tilde{\nu}_{\lambda}^{(0)} (   \tilde{\epsilon} ) & = &
 \Theta ( \tilde{\epsilon} ) 
 \left[  \frac{\sqrt{ \tilde{\epsilon}}}{\pi} - 
  \frac{\lambda}{2} + \frac{1}{\pi} \arctan \left( \frac{1}{ \sqrt{\tilde{\epsilon}} } \right) 
 \right]
 \nonumber
 \\
 &+ & \Theta ( -\tilde{\epsilon} )   \delta_{ \lambda , -1}    \Theta ( 1 - \sqrt{-\tilde{\epsilon}} ) ,
 \\
  \tilde{\nu}_{\lambda}^{(2)} (   \tilde{\epsilon} ) & = &
 - \frac{\Theta ( \tilde{\epsilon} )}{2} \frac{  \sqrt{  \tilde{\epsilon}}}{ \pi ( 1 + \tilde{\epsilon} )^2} 
 \nonumber
 \\
 & + &  \frac{\Theta ( -\tilde{\epsilon} )}{4}   \delta_{ \lambda , -1}  
 \delta^{\prime} ( 1 - \sqrt{ - \tilde{\epsilon} } ),
 \end{eqnarray}
and  $\delta^{\prime} (x) = \frac{d}{dx} \delta ( x )$ is the derivative of the $\delta$-function with respect to its argument. The order parameter expansion of the weighted density of states is
\begin{equation}
 \frac{\sigma_{\lambda} ( \epsilon, p )}{\nu_0} = \tilde{\sigma}_{\lambda}^{(0)} 
 \left( \frac{ \epsilon - \Sigma_0 }{\epsilon_0} \right)
   +   \tilde{p}^2  \tilde{\sigma}_{\lambda}^{(2)} 
  \left( \frac{ \epsilon - \Sigma_0 }{\epsilon_0} \right) 
  + {\cal{O}} ( \tilde{p}^4 ),
 \label{eq:sigmaexpansion}
 \end{equation}
with
 \begin{eqnarray}
 \tilde{\sigma}_{\lambda}^{(0)} (   \tilde{\epsilon} ) & = &
 \frac{\Theta ( \tilde{\epsilon} )}{4}
 \left\{
 ( 3 + \tilde{\epsilon} )  \left[
 \frac{ 1 }{2}  -   \frac{ \lambda   }{\pi}   \arctan \left( \frac{1}{ \sqrt{\tilde{\epsilon}} } \right) 
 \right]     -  \lambda    \frac{3 \sqrt{ \tilde{\epsilon}}}{  \pi}    \right\}
 \nonumber
 \\
 &+ & \frac{\Theta ( -\tilde{\epsilon} )}{4}   \delta_{ \lambda , -1}    \Theta ( 1 - \sqrt{-\tilde{\epsilon}} ) 
 ( 3 + \tilde{\epsilon} ) ,
 \\
  \tilde{\sigma}_{\lambda}^{(2)} (   \tilde{\epsilon} ) & = &
  \frac{\Theta (  \tilde{\epsilon} )}{16} 
 \left[ \frac{  \lambda \sqrt{  \tilde{\epsilon}} ( 1 - \tilde{\epsilon})}{ \pi  
 ( 1 + \tilde{\epsilon} )^2}      - \frac{1}{2}     +
 \frac{ \lambda   }{\pi}   \arctan \left( \frac{1}{ \sqrt{\tilde{\epsilon}} } \right) 
 \right]
 \nonumber
 \\
 & + &  \frac{\Theta ( -\tilde{\epsilon} )}{16}   \delta_{ \lambda , -1}  
 \Bigl[- \Theta  ( 1 - \sqrt{ - \tilde{\epsilon} } )      
+ 2 \delta  ( 1 - \sqrt{ - \tilde{\epsilon} } )
 \nonumber
 \\
 & &
    \hspace{18mm} + \frac{3 + \tilde{\epsilon}}{2}
 \delta^{\prime} ( 1 - \sqrt{ - \tilde{\epsilon} } )
  \Bigr].
 \end{eqnarray}
At low temperatures only the leading asymptotics of these functions
close to the bottom of the lower energy branch is relevant.
Shifting the dimensionless energy as $\tilde{\epsilon} = - 1 + \omega$, we may approximate
for $ |\omega |   \ll 1$,
 \begin{eqnarray}
 \tilde{\nu}_-^{(0)} ( -1 + \omega ) & \approx & \Theta ( \omega ),
 \label{eq:nu0}
 \\
 \tilde{\sigma}_-^{(0)} ( -1 + \omega ) & \approx & \frac{  \Theta ( \omega )}{2},
 \label{eq:sigma0}
\\
 \tilde{\nu}_-^{(2)} ( -1 + \omega ) & \approx & \frac{1}{4} \delta^{\prime} 
 (  1 - \sqrt{1 - \omega } ) \approx
  \delta^{\prime} ( \omega ),
 \\
\tilde{\sigma}_-^{(2)} ( -1 + \omega ) & \approx & \frac{1}{16}
 \left[ - \Theta ( \omega ) + 2 \delta ( \omega ) + 4  \delta^{\prime} ( \omega ) \right],
 \label{eq:sigma2}
 \hspace{7mm}
 \end{eqnarray}
where we have used $\delta^{\prime} ( a \omega ) = a^{-2} 
\delta^{\prime} ( \omega )$ and $\omega \delta^{\prime} ( \omega ) = - \delta ( \omega )$.

\section*{APPENDIX C: Order parameter expansion}

\renewcommand{\theequation}{C\arabic{equation}}
\setcounter{equation}{0}

For temperatures slightly below the critical temperature
the order parameter $\tilde{p} = | \tilde{g}_{\bot} | \tilde{M} / 2$ is small
so that we may expand all quantities to second order in powers of $\tilde{p}$.
Having expressed all quantities in terms of the
spectral densities ${\nu}_{\lambda} ( \epsilon, p )$ and
$\sigma_{\lambda} ( \epsilon , p )$, we use the order parameter expansion
of these, given in Eqs.~(\ref{eq:nuexpansion}, \ref{eq:sigmaexpansion}),
to generate the corresponding expansion of any other quantity.
Using the density equation (\ref{eq:rhotildedef})
we obtain for the chemical potential
 \begin{equation}
 \mu = \mu_0 + \mu_2 \tilde{p}^2 + {\cal{O}} ( \tilde{p}^4 ),
 \end{equation}
where the chemical potential for vanishing order parameter
is determined by
 \begin{equation}
 \tilde{\rho} =  \int_{- \infty}^{\infty} d \tilde{\epsilon} \sum_{\lambda} \tilde{\nu}^{(0)}_{\lambda}
 ( \tilde{\epsilon} ) n_0 ( \tilde{\epsilon} ),
 \label{eq:zerothdens}
 \end{equation}
with
\begin{equation}
  n_0 ( \tilde{\epsilon} ) = \frac{1}{ e^{ ( \tilde{\epsilon} + \tilde{\Sigma}_0 - \tilde{\mu}_0 ) / \tau } -1 }.
 \end{equation}
Here $\tilde{\Sigma}_0 = \Sigma_0 / \epsilon_0$ and
$\tilde{\mu}_0 = \mu_0 / \epsilon_0$.
The leading correction to the chemical potential in the symmetry broken phase is
 \begin{equation}
 \frac{ \mu_2 }{T} = \frac{ \tilde{\mu}_2 }{\tau }
 = 
 \frac{ \int_{- \infty}^{\infty} d \tilde{\epsilon} \sum_{\lambda} \tilde{\nu}^{(2)}_{\lambda}
 ( \tilde{\epsilon} ) n_0 ( \tilde{\epsilon} ) }{
   \int_{- \infty}^{\infty} d \tilde{\epsilon} \sum_{\lambda} \tilde{\nu}^{(0)}_{\lambda}
 ( \tilde{\epsilon} ) n_0 ( \tilde{\epsilon} ) [ n_0 ( \tilde{\epsilon} ) +1 ]      }.
 \end{equation}
The corresponding expansion of the  susceptibility $\tilde{\chi}_{\bot}$
defined in Eq.~(\ref{eq:tildechi}) is
 \begin{equation}
 \tilde{\chi}_{\bot} = \tilde{\chi}_{\bot}^{(0)} + \tilde{p}^2  \tilde{\chi}_{\bot }^{( 2) }
 +  {\cal{O}} ( \tilde{p}^4 ),
 \end{equation}
with
 \begin{equation}
\tilde{\chi}_{\bot}^{(0)} = - \frac{1}{2 \tau} 
 \int_{- \infty}^{\infty} d \tilde{\epsilon} \sum_{\lambda} \lambda \tilde{\sigma}^{(0)}_{\lambda}
 ( \tilde{\epsilon} ) n_0 ( \tilde{\epsilon} ) [ n_0 ( \tilde{\epsilon} ) +1 ]   ,
 \label{eq:chizerodef}
 \end{equation}
and
 \begin{eqnarray}
\tilde{\chi}_{\bot}^{(2)} & = & - \frac{1}{2 \tau}
 \int_{- \infty}^{\infty} d \tilde{\epsilon} \sum_{\lambda} \lambda 
 \biggl\{ \tilde{\sigma}^{(2)}_{\lambda}   ( \tilde{\epsilon} )
 n_0 ( \tilde{\epsilon} ) [ n_0 ( \tilde{\epsilon} ) +1 ]
  \nonumber
 \\
 &- &
      \frac{ \tilde{\mu}_2}{\tau }  \tilde{\sigma}^{(0)}_{\lambda}
  ( \tilde{\epsilon} )
  n_0 ( \tilde{\epsilon} ) [ n_0 ( \tilde{\epsilon} ) +1 ]    [  2 n_0 ( \tilde{\epsilon} ) +1 ] \biggr\}.
 \end{eqnarray}
And finally, the expansion of the dimensionless free energy defined in Eq.~(\ref{eq:freedef}) is
  \begin{equation}
 f \equiv \frac{ \Omega + \mu N}{ V \nu_0 \epsilon_0^2}
 = f_0 + \tilde{p}^2 f_2  +  {\cal{O}} ( \tilde{p}^4 ),
 \end{equation}
with
  \begin{eqnarray}
 f_0 & = & \tau \int_{- \infty}^{\infty} d \tilde{\epsilon}
 \sum_{\lambda} \tilde{\nu}^{(0)}_{\lambda} (  \tilde{\epsilon} )
 \ln \left[ 1 -  e^{ - ( \tilde{\epsilon} + \tilde{\Sigma}_0 - \tilde{\mu}_0) / \tau } \right] 
 \nonumber
 \\
 & - &  \frac{1}{2} \left[  \tilde{g}_{\parallel}  + \frac{\tilde{g}_{\bot}}{2} 
 \right]  \tilde{\rho}^2 +  \tilde{\mu}_0 \tilde{\rho},
 \\
 f_2 & = &  \frac{4}{-g_{\bot}} +
 \tau \int_{- \infty}^{\infty} d \tilde{\epsilon}
 \sum_{\lambda} \tilde{\nu}^{(2)}_{\lambda} (  \tilde{\epsilon} )
 \ln \left[ 1 -  e^{ - ( \tilde{\epsilon} + \tilde{\Sigma}_0 - \tilde{\mu}_0) / \tau } \right] .
 \nonumber
 \\
 & &
 \end{eqnarray}

In the low-temperature regime $ \tau \ll 1$ the above expressions can be evaluated
analytically because it is then allowed to substitute
the leading asymptotics of the spectral densities
for energies close to the bottom of the
lower branch given in Eqs.~(\ref{eq:nu0}--\ref{eq:sigma2}).
The zeroth order chemical potential and the
corresponding fugacity $z_0$ are then given by
 \begin{equation}
 z_0 = e^{ ( \tilde{\mu}_0 - \tilde{\Sigma}_0 +1 ) / \tau } = 1 - e^{ - \tilde{\rho} / \tau },
 \end{equation}
while the correction to the chemical potential is
\begin{equation}
 \tilde{\mu}_2 = \frac{1}{\tau ( 1 -z_0)} = \frac{e^{\tilde{\rho} / \tau}}{\tau }.
 \end{equation}
The zeroth order susceptibility
 $\tilde{\chi}_{\bot}^{(0)}$ is given by the right-hand side
of Eq.~(\ref{eq:chilow}), while the leading correction is
\begin{equation}
 \tilde{\chi}_{\bot}^{(2)} \approx - \frac{z_0 ( 3 + z_0)}{8 \tau^2 (1 - z_0)^3}.
 \end{equation}
The  order parameter close to the critical point shows the usual mean-field behavior,
 \begin{equation}
 \tilde{p} \sim (1 - z_0 ) \sqrt{ \frac{ 2 \tilde{\rho} \tau_c}{z_0 ( 3 + z_0)} } 
 \sqrt{ \frac{ \tau_c - \tau }{\tau_c}}.
 \end{equation}
At the critical temperature we may write
 \begin{equation}
 1 - z_0 = e^{ - \tilde{\rho}/ \tau_c} = \frac{ 1}{1+ \frac{4 }{ | \tilde{g}_{\bot} |} },
 \end{equation}
so that we arrive at Eq.~(\ref{eq:Mcrit}) for 
the transverse magnetization $\tilde{M} = 2 \tilde{p} / | \tilde{g}_{\bot} | $.
Finally, the leading coefficient in the expansion of the free energy is
 \begin{equation}
 f_0 = - \tau^2 {\rm Li}_2 ( z_0 )
-\frac{1}{2} \left[  \tilde{g}_{\parallel}  + \frac{\tilde{g}_{\bot}}{2} 
 \right]  \tilde{\rho}^2 +  \tilde{\mu}_0 \tilde{\rho},
 \end{equation}
where ${\rm Li}_2 ( z_0 )$ is the polylogarithm.
The coefficient of the term proportional to $\tilde{p}^2$ is
\begin{equation}
 f_2 =  -  \frac{4}{ \tilde{g}_{\bot}}   - \frac{ z_0}{1-z_0}  =
 4 \left[  \frac{1}{ | \tilde{g}_{\bot} |} - \tilde{\chi}_{\bot}^{(0)} \right].
 \end{equation}
Note that for $\tau < \tau_c$ the right-hand side of this expression is negative,
so that in the ferromagnetic phase the system indeed gains energy.

\section*{APPENDIX D: Pair condensation}

\renewcommand{\theequation}{D\arabic{equation}}
\setcounter{equation}{0}

For attractive $g_\bot$ the system can also exhibit a pairing instability 
which competes with the ferromagnetic
instability discussed in the main text. In this appendix we show however that, 
at least for sufficiently low densities,
the ferromagnetic instability is dominant. 

\begin{figure}[tb]    
  \centering
  \includegraphics[keepaspectratio,width=\linewidth]{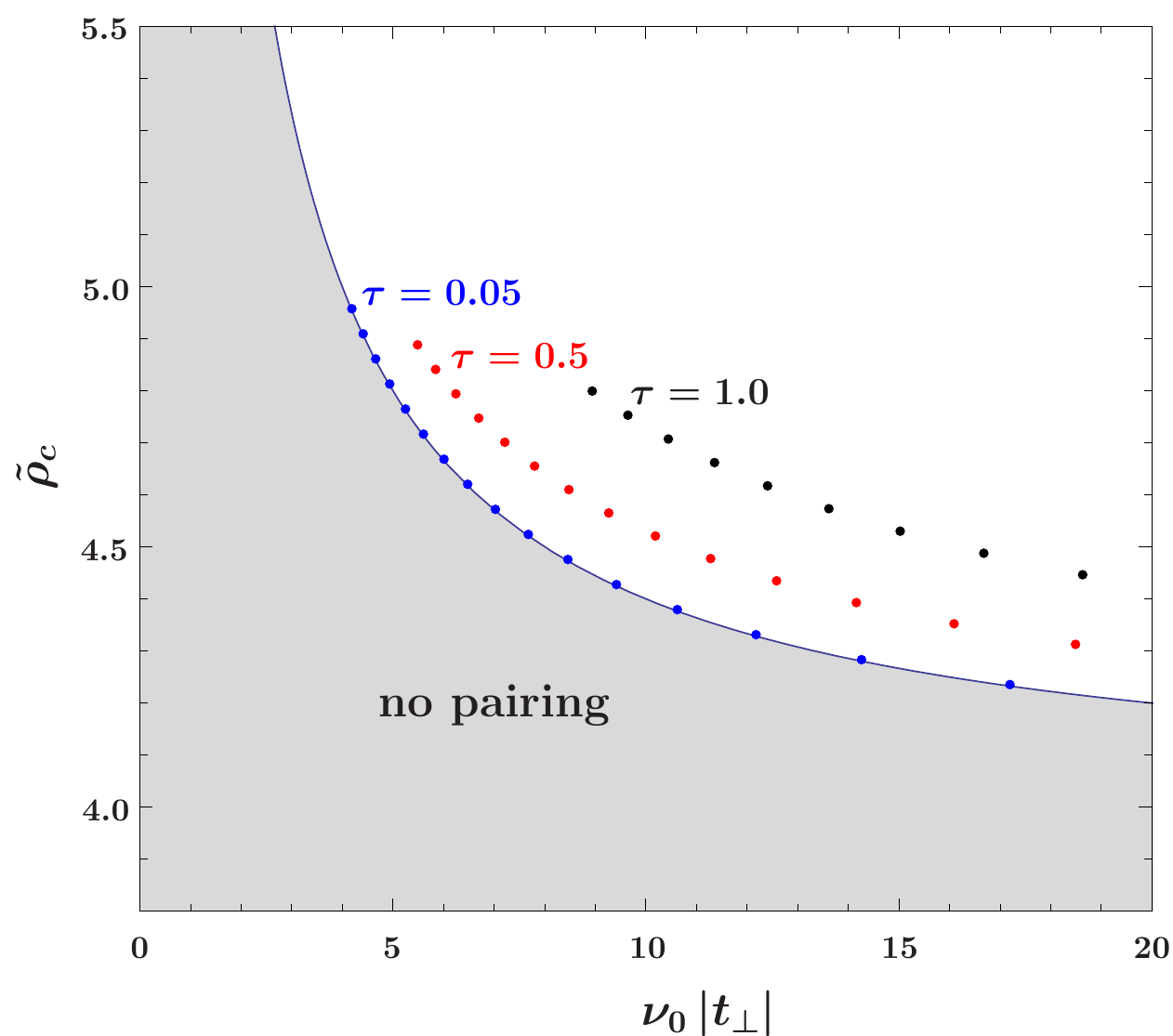}  
  \caption{\label{fig:9}(Color online) %
Critical density $\tilde{\rho}_c$ for pair condensation as a function of the two-body $t$-matrix 
$| t_{\bot} |$ (note that $t_\bot < 0$, so $-t_\bot=|t_\bot|$) for fixed temperatures $\tau$. The
solid line depicts $\tilde{\rho}_c$ for vanishing temperature as
given in Eq.~(\ref{limiting}). The dots represent numerical solutions of the regularized gap equation 
(\ref{eq:selfconpair}) for the indicated temperatures. The shaded area underneath the solid line 
represents the regime where pairing cannot occur at any temperature.
}  
 \label{fig:nopairing}
\end{figure}

If we decouple the interaction in the particle-particle channel, we obtain the
mean-field Hamiltonian
 \begin{eqnarray}
 {\cal{H}}_{\rm MF} - \mu {\cal{N}} & = & 
 \sum_{\bd{k}} \Bigl[ ( E_{\bd{k} -} - \mu )  a^{\dagger}_{\bd{k}} a_{\bd{k}} 
 + ( E_{\bd{k} +} - \mu )  b^{\dagger}_{\bd{k}} b_{\bd{k}} 
 \nonumber
 \\
 & & \hspace{-20mm} + \bar{\chi}_0 e^{ i \varphi_{\bd{k}} } a_{- \bd{k}} b_{\bd{k}} +
 \chi_0 e^{ - i \varphi_{\bd{k}} } b^{\dagger}_{ \bd{k}} a^{\dagger}_{-\bd{k}}
 \Bigr] + \frac{V}{| g_{\bot} |} | \chi_0 |^2.
 \end{eqnarray}
In general, the boson-pairing order parameter $\chi_0 = | \chi_0 | e^{i \alpha}$ is complex,
but all phases can be eliminated by setting $\tilde{a}_{\bd{k}} = e^{ i \varphi_{\bd{k}} - i \alpha} a_{\bd{k}}$.
The Hamiltonian can then be diagonalized by means of a Bogoliubov transformation,
 \begin{equation}
 \left( \begin{array}{c} \tilde{a}_{\bd{k}} \\ b_{\bd{k}}^{\dagger}  \end{array}
 \right) = \left( \begin{array}{cc} u_{\bd{k}}^{\ast} & - v_{\bd{k}}^{\ast} \\
 - v_{\bd{k}} & u_{\bd{k}} \end{array} \right) 
 \left( \begin{array}{c} {\alpha}_{\bd{k}} \\ \beta_{\bd{k}}^{\dagger}  \end{array}
 \right),
 \end{equation}
with
 \begin{eqnarray}
 u_{\bd{k}} & = & 
 \sqrt{ \frac{ \xi_{\bd{k}} + \omega_{\bd{k}} }{2 \omega_{\bd{k}} }},
 \; \; \; 
  v_{\bd{k}}  =  
 \sqrt{ \frac{ \xi_{\bd{k}} - \omega_{\bd{k}} }{2 \omega_{\bd{k}} }},
 \end{eqnarray}
where
 \begin{eqnarray}
 \xi_{\bd{k}} & = & \frac{ \bd{k}^2}{2m} - \mu ,
 \; \; \; 
 \omega_{\bd{k}}  =  \sqrt{ \xi_{\bd{k}}^2 - | \chi_0 |^2 }.
 \end{eqnarray}
The corresponding grand canonical potential is
 \begin{eqnarray}
 \Omega_{\rm MF} & = & T \sum_{\bd{k} \lambda} \ln \left[
 1 - e^{ - \beta ( \omega_{\bd{k}} + \lambda v_0 | \bd{k}_{\bot} | ) } \right]
 \nonumber
 \\
 & + & \sum_{\bd{k}} ( \omega_{\bd{k}} - \xi_{\bd{k}} ) + \frac{V}{| g_{\bot} |} | \chi_0 |^2.
 \end{eqnarray}
The self-consistency equation for the order parameter is
 \begin{equation}
 \frac{1}{ | g_{\bot} | } = \frac{1}{2V} \sum_{\bd{k} \lambda}  \frac{1}{\omega_{\bd{k}} }
 \left[
 \frac{1}{ e^{  \beta ( \omega_{\bd{k}} + \lambda v_0 | \bd{k}_{\bot} | ) } -1 } + \frac{1}{2} \right],
 \end{equation}
while the density equation is
  \begin{equation}
 \rho = \frac{1}{V} \sum_{\bd{k} \lambda}  \frac{\xi_{\bd{k}}}{\omega_{\bd{k}} }
 \frac{1}{ e^{  \beta ( \omega_{\bd{k}} + \lambda v_0 | \bd{k}_{\bot} | ) } -1 }
 + \frac{1}{V} \sum_{\bd{k}} \left[  \frac{\xi_{\bd{k}}}{\omega_{\bd{k}} } -1 \right].
 \end{equation}
The integrals are ultraviolet divergent and must be regularized. One possibility is to
eliminate the bare interaction in favor of the two-body 
t-matrix ${t}_{\bot}$,
which can be defined via
 \begin{equation}
  \frac{1}{  t_{\bot}  } = \frac{1}{{g}_{\bot}} + \frac{1}{V} \sum_{\bd{k}} \frac{1}{ 2 \epsilon_{\bd{k}}},
 \end{equation}
where $\epsilon_{\bd{k}}=\bd{k}^2/2m$. The regularized gap equation is then
 \begin{eqnarray}
 & & \frac{1}{ - t_{\bot}  } + \frac{1}{2V} \sum_{\bd{k}} \left( 
 \frac{1}{\epsilon_{\bd{k}}} - \frac{1}{\omega_{\bd{k}}}  \right) 
 \nonumber
 \\
& = &  \frac{1}{2V} \sum_{\bd{k} \lambda}  \frac{1}{\omega_{\bd{k}} }
 \frac{1}{ e^{  \beta ( \omega_{\bd{k}} + \lambda v_0 | \bd{k}_{\bot} | ) } -1 }.
 \label{eq:selfconpair}
 \end{eqnarray}
At the critical point we set $\chi_0 =0$. 
At low temperatures,  we may replace
$\xi_{\bd{k}} \approx \epsilon_0 - \mu = 2 \epsilon_0$ 
in the prefactor on the right-hand side.
Then we obtain for the critical density $\tilde{\rho}_c =  \rho_c / ( \epsilon_0 \nu_0)$
at zero temperature
 \begin{equation}
 \tilde{\rho}_c = 4 \left[ \frac{1}{ -(\nu_0 t_\bot )}  + 1 \right]. 
 \label{limiting}
 \end{equation}
For densities smaller than $\tilde{\rho}_c$ 
there is no pairing instability at zero temperature.
For finite temperatures we have solved the regularized gap equation 
(\ref{eq:selfconpair}) numerically, see Fig.~\ref{fig:nopairing}. 
Because the zero-temperature result (\ref{limiting}) for the critical density forms
a lower bound for the critical density at finite temperatures,
we conclude that for sufficiently small densities, pairing
cannot occur at any temperature. For densities smaller than this threshold the ferromagnetic instability discussed in the main text is dominant.

\end{appendix}

\end{document}